\documentclass[iop]{emulateapj}

\newcommand{\kms}{\textrm{km~s$^{-1}$}}

\newcommand{\ergs}{\textrm{erg~s$^{-1}$}}
\newcommand{\lsolar}{L$_{\odot}$}
\newcommand{\msolar}{M$_{\odot}$}
\newcommand{\ml}{M$_{\odot}$ yr$^{-1}$}
\newcommand{\mdot}{\dot{M}}

\usepackage{subfigure}
\usepackage{multirow}
\usepackage[dvips]{color}
\definecolor{Mygrey}{gray}{0.6}

\begin{document}

\title{A $Spitzer$~Survey for Dust in Type IIn Supernovae}
\shorttitle{$Spitzer$~Type IIn Survey}
\author{Ori D. Fox\altaffilmark{1}, Roger A. Chevalier\altaffilmark{2}, Michael F. Skrutskie\altaffilmark{2}, Alicia M. Soderberg\altaffilmark{3}, Alexei V. Filippenko\altaffilmark{4}, Mohan Ganeshalingam\altaffilmark{4}, Jeffrey M. Silverman\altaffilmark{4}, Nathan Smith\altaffilmark{4,5}, and Thea N. Steele\altaffilmark{4}}
\altaffiltext{1}{NASA Goddard Space Flight Center, Greenbelt, MD 20771.}
\altaffiltext{2}{Department of Astronomy, University of Virginia, Charlottesville, VA 22903.}
\altaffiltext{3}{Harvard-Smithsonian Center for Astrophysics, Cambridge, MA 02138.}
\altaffiltext{4}{Department of Astronomy, University of California, Berkeley, CA 94720-3411.}
\altaffiltext{5}{Department of Astronomy, Steward Observatory, Tucson, AZ 85721}
\email{ori.d.fox@nasa.gov .}

\begin{abstract}

Recent observations suggest that Type IIn supernovae (SNe~IIn) may exhibit late-time ($>$100 days) infrared (IR) emission from warm dust more than other types of core-collapse SNe.  Mid-IR observations, which span the peak of the thermal spectral energy distribution, provide useful constraints on the properties of the dust and, ultimately, the circumstellar environment, explosion mechanism, and progenitor system.  Due to the low SN~IIn rate ($<$10\% of all core-collapse SNe), few IR observations exist for this subclass.  The handful of isolated studies, however, show late-time IR emission from warm dust that, in some cases, extends for five or six years post-discovery.  While previous $Spitzer$/IRAC surveys have searched for dust in SNe, none have targeted the Type IIn subclass.  This article presents results from a warm $Spitzer$/IRAC survey of the positions of all 68 known SNe~IIn within a distance of 250~Mpc between 1999 and 2008 that have remained unobserved by $Spitzer$~more than 100 days post-discovery.  The detection of late-time emission from ten targets ($\sim$15\%) nearly doubles the database of existing mid-IR observations of SNe~IIn.  Although optical spectra show evidence for new dust formation in some cases, the data show that in most cases the likely origin of the mid-IR emission is pre-existing dust, which is continuously heated by optical emission generated by ongoing circumstellar interaction between the forward shock and circumstellar medium.  Furthermore, an emerging trend suggests that these SNe decline at $\sim$1000--2000 days post-discovery once the forward shock overruns the dust shell.  The mass-loss rates associated with these dust shells are consistent with luminous blue variable (LBV) progenitors.

\end{abstract}

\keywords{circumstellar matter --- supernovae: general --- supernovae: individual --- dust, extinction --- infrared: stars}

\clearpage

\section{Introduction}
\label{sec:intro}

Type IIn supernovae (SNe; see \citealt{filippenko97} for a review) have gained considerable attention over the past decade.  Representing {\it less} than 10\% of all core-collapse events \citep{smartt09,smith11rates,li11}, this subclass comprises {\it more} than 50\% of the rare sample of SNe observed to exhibit a late-time ($>$100 days) infrared (IR) excess with respect to the optical blackbody counterpart, typically suggesting the presence of warm dust \citep[e.g.,][and see Table \ref{tab1}]{gerardy02,fox09,fox10}.  Named for their relatively ``narrow'' ($\lesssim 700$ km s$^{-1}$) emission lines arising from a dense, slowly moving, pre-existing circumstellar medium (CSM) \citep{schlegel90}, these SNe~IIn may be more prone to have warm dust than any other SN type since (1) dust is able to form in the cool, dense, post-shock layers not available in other SNe \citep[e.g.,][]{pozzo04,smith09ip}, and (2) the dense CSM produced by pre-SN mass loss can contain dust that is illuminated by the supernova radiation \citep[e.g.,][]{smith08gy,miller10gy}.

\begin{deluxetable*}{ l c c c}
\tablewidth{0pt}
\tabletypesize{\footnotesize}
\tablecaption{All SNe with {\bf Observed} Late-Time IR Excess \label{tab1}}
\tablecolumns{4}
\tablehead{
\multirow{2}*{SN} & \multirow{2}*{Subclass} & \colhead{$Spitzer$~Observations?/} & \multirow{2}*{Reference}\\
\colhead{} & \colhead{} & \colhead{Detections?} & \colhead{}
}
\startdata
1979c & II & No/No  & \citet{merrill80}\\
1980k & II & No/No & \citet{telesco81}\\
1982e & ? & No/No  & \citet{graham83,graham86}\\
1982l & II & No/No  & \citet{graham86}\\
1982r & I & No/No  & \citet{graham86}\\
1985l & II & No/No  & \citet{elias86,fesen98}\\
1986J & IIn & No/No  & \citet{weiler90,leibundgut91}\\ 
1987A & II & Yes/Yes  & Many Authors\\
1993J & IIb & No/No  & \citet{lewis94,matheson00}\\
1994Y & ? & No/No  & \citet{garnavich96}\\
1995G & IIn & No/No  & \citet{pastorello02}\\
1995N & IIn & No/No  & \citet{gerardy02}\\
1997ab & IIn & No/No  & \citet{gerardy02}\\
1998S & IIn & No/No  & \citet{gerardy02,pozzo04}\\
1999Z & IIn & No/No  & \citet{gerardy02}\\
1999bw & IIn & Yes/Yes  & \citet{sugerman04}\\
1999el & IIn & No/No  & \citet{gerardy02,carlo02}\\
1999em & IIP & No/No  & \citet{elmhamdi03}\\
2002bu & IIn & Yes/Yes  & \citet{thompson09}\\
2002hh & IIP & Yes/Yes  & \citet{meikle06}\\
2002ic & Ia/IIn & No/No  & \citet{kotak04}\\
2003gd & IIP & Yes/Yes  & \citet{sugerman06,meikle07}\\
2004dj & IIP & Yes/Yes  & \citet{kotak05,kotak06,szalai11}\\
2004et & IIP & Yes/Yes  & \citet{kotak09,maguire10et}\\
2005af & IIP & No/No  & \citet{kotak06}\\
2005kd & IIn & No/No  & \citet{tsvetkov08}\\
2005ip & IIn & No/No  & \citet{fox09,smith09ip}\\
2006gy & IIn & No/No  & \citet{smith09gy}\\
2006jc & Ib & No/No  & \citet{smith08jc}\\
2006tf & IIn & No/No  & \citet{smith08tf}\\
2007it & IIP & Yes/Yes & \citet{andrews11it}\\
2007od & IIP & No/No  & \citet{andrews10}\\
2007rt & IIn & No/No & \citet{trundle09}\\
2008iy & IIn & No/No & \citet{miller10iy}
\enddata
\end{deluxetable*}

The origin and heating mechanism of the dust, however, can be ambiguous.  Disentangling the various dust models offers important clues regarding the supernova's circumstellar environment, explosion mechanism, and even progenitor system \citep[e.g.,][]{fox10}.  For example, if shock heated, the dust temperature correlates with the gas density \citep{draine79,draine81,dwek87,dwek08}, which can be used to trace the progenitor's mass-loss history \citep[e.g.,][]{smith09ip,fox10}.  Alternatively, the discovery of significant amounts of newly formed SN dust would provide the much sought evidence necessary to confirm SN dust models \citep[e.g.,][]{nozawa03,nozawa08}.  At this point, the observed dust yields all tend to be 2--3 orders of magnitude smaller than predicted by the models \citep[e.g.,][and references therein]{kozasa09,meikle11}, although far-IR observations of SN 1987A suggest larger reservoirs of dust may be hidden at colder temperatures \citep[$\sim$20 K,][]{matsuura11}.

Mid-IR observations, which span the peak of the thermal emission from warm dust, offer the most useful constraints on the dust properties.  Yet observations at these wavelengths remain sparse.  In fact, the \citet{fox10} {\it Spitzer Space Telescope} IRS spectrum of SN 2005ip is the {\it only} mid-IR spectrum of a SN~IIn to date.  Estimated to have a rate of no more than 10 yr$^{-1}$ out to 150~Mpc \citep{dahlen99}, the number of nearby SNe~IIn is small, and SNe~IIn at greater distances are not easy targets.

The few observed dust-emitting SNe~IIn, however, are anomalously luminous and linger for many years at mid-IR wavelengths, thereby making remnant archeology possible.  While 79 SNe~IIn within a distance of 250~Mpc were discovered during the years 1999--2008, only 11 of these positions have archived $Spitzer$ IRAC data collected (either intentionally or serendipitously) more than a few months post-discovery.  Five of these eleven events ($\sim$45\%) show evidence for late-time IR emission, in some cases up to six years post-discovery: SNe 1999bw \citep{sugerman04,thompson09}, 1999el \citep{carlo02}, 2002bu \citep{thompson09}\footnote{Note that \citet{smith11bu} suggest that SNe 1999bw and 2002bu may be SN impostors instead of genuine SNe.}, 2003lo \citep{meikle05}, and 2005ip \citep{fox09,fox10}.

We therefore executed a warm $Spitzer$~follow-up survey of the remaining 68 Type IIn events discovered in the years 1999--2008.\footnote{We also included the well-studied Type IIn SN 1997ab, for a total of 69 SNe.}  This paper presents the results of the survey, including the positive detection of late-time emission from ten targets, which more than doubles the database of existing mid-IR observations of SNe~IIn.  For some targets, optical and near-IR photometry and spectroscopy also exist. In \S \ref{sec:obs} we summarize the observations and data-reduction techniques. $Spitzer$ photometry constrains the dust mass, temperature, and, thereby, luminosity. Section \ref{sec:source} implements the methods presented by \citet{fox10} to explore the origin and heating mechanism of these components; in most cases, results are compared to those of SN 2005ip.  Section \ref{sec:model} discusses the overall trends, circumstellar model, and progenitor system, while \S \ref{sec:con} summarizes the results and discusses future work.

\section{Observations}
\label{sec:obs}

\subsection{Mid-Infrared Warm $Spitzer$~IRAC Survey and Photometry}
\label{sec:irac}

\begin{deluxetable*}{ l c c c c c c }
\tablewidth{0pt}
\tabletypesize{\normalsize}
\tablecaption{$Spitzer$~Survey Targets \label{tab2}}
\tablecolumns{7}
\tablehead{
\colhead{SN} & \colhead{JD} & \colhead{Epoch} & \colhead{$\alpha$ (h:m:s)} & \colhead{$\delta$ ($^\circ$:$'$:$''$)} & \colhead{Distance  \tablenotemark{*}} & \colhead{$t_{\rm int}$}\\
\colhead{}&\colhead{($-$2,450,000)}&\colhead{(days)}&\colhead{(J2000)}&\colhead{(J2000)}&\colhead{(Mpc)}&\colhead{(s)}
}
\startdata
1997ab & 5140& 5018& 09:51:00.40 & +20:04:24.0 & 50 & 300\\
1999Z & 5065& 4104& 10:22:37.23 & +27:21:19.8 & 210 & 1800 \\
1999eb & 5170& 3612& 01:43:45.45 & +04:13:25.9 & 75 & 300 \\
2000P & 5322& 3450& 13:07:10.53 & $-$28:14:02.5 & 31 & 300\\
2000cl & 5063& 3631& 10:37:16.07 & $-$41:37:47.8 & 38 & 300 \\
2000ct & 5066& 3335& 17:01:03.64 & +33:28:45.0 & 124 & 300 \\
2000eo & 5068& 3247& 03:09:08.17 & $-$10:17:55.3 & 44 & 300 \\
2000ev & 5174& 3264& 06:47:52.00 & +84:10:02.2 & 61 & 300 \\
2001I & 5322& 3180& 03:43:57.28 & +39:17:39.4 & 68 & 300 \\
2001ey & 5074& 3002& 22:26:30.77 & $-$06:23:28.6 & 105 & 300 \\
2001fa & 5065& 2872& 01:48:22.22 & +11:31:34.4 & 71 & 300 \\
2001ir & 5209& 2937& 08:36:28.12 & $-$11:50:03.5 & 81 & 300 \\
2002A & 5075& 2864& 07:22:36.14 & +71:35:41.5 & 40 & 300 \\
2002bj & 5209& 2784& 05:11:46.41 & $-$15:08:10.8 & 50 & 300 \\
2002bv & 5146& 2796& 07:50:01.13 & +30:01:32.2 & 115 & 300 \\
2002cb & 5186& 2950& 13:04:23.97 & +47:35:53.2 & 123 & 300 \\
2002ea & 5118& 2584& 02:08:25.08 & +14:20:52.8 & 61 & 300 \\
2002fj & 5108& 2792& 08:40:45.10 & $-$04:07:38.5 & 60 & 300 \\
2003as & 5078& 2455& 05:28:45.81 & +49:52:59.1 & 95 & 300 \\
2003ei & 5095& 2296& 16:28:40.30 & +12:46:05.8 & 110 & 300 \\
2003hy & 5084& 2290& 21:54:22.72 & +15:09:37.8 & 101 & 300 \\
2003hy & 5084& 2290& 21:54:22.72 & +15:09:37.8 & 101 & 300 \\
2003jh & 5202& 2166& 04:16:37.30 & $-$12:23:32.9 & 122 & 300 \\
2004F & 5151& 2067& 03:17:53.80 & $-$07:17:43.0 & 73 & 300 \\
2004ay & 5065& 1977& 18:28:57.57 & +51:38:55.7 & 133 & 300 \\
2004ec & 5322& 1825& 17:08:04.96 & +26:22:44.4 & 171 & 900 \\
2004gd & 5113& 1835& 07:09:11.71 & +20:36:10.6 & 72 & 300 \\
2005R & 5075& 1938& 11:14:56.81 & +33:49:35.7 & 151 & 600 \\
2005aq & 5135& 1671& 04:31:38.82 & $-$04:35:06.8 & 56 & 300 \\
2005av & 5066& 1657& 20:44:37.58 & $-$68:45:10.6 & 43 & 300 \\
2005bx & 5065& 1711& 13:50:24.95 & +68:33:19.4 & 132 & 300 \\
2005cl & 5084& 1635& 21:02:02.35 & $-$06:17:35.7 & 107 & 300 \\
2005cp & 5066& 1523& 23:59:30.88 & +18:12:09.6 & 91 & 300 \\
2005cy & 5066& 1498& 18:26:49.11 & +51:08:30.4 & 138 & 300 \\
2005db & 5111& 1495& 00:41:26.79 & +25:29:51.6 & 62 & 300 \\
2005dr & 5108& 1476& 05:02:17.18 & +07:38:21.4 & 157 & 600 \\
2005ds & 5322& 1470& 17:57:39.89 & +27:50:18.4 & 97 & 300 \\
2005ep & 5151& 1421& 16:56:35.49 & +58:01:30.2 & 121 & 300 \\
2005gl & 5072& 1417& 00:49:50.02 & +32:16:56.8 & 65 & 300 \\
2005gn & 5065& 1479& 05:48:49.07 & $-$24:22:45.5 & 165 & 600 \\
2005ly & 5108& 1349& 01:23:28.84 & +30:46:45.7 & 143 & 600 \\
2005ma & 5187& 1384& 04:49:53.91 & $-$10:45:23.4 & 62 & 300 \\
2006aa & 5075& 1547& 11:53:19.89 & +20:45:18.2 & 85 & 300 \\
2006am & 5139& 1276& 14:27:37.24 & +41:15:35.4 & 37 & 300 \\
2006bo & 5322& 1320& 20:30:41.90 & +09:11:40.8 & 63 & 300 \\
2006bv & 5077& 1348& 12:41:01.55 & +63:31:11.6 & 35 & 300 \\
2006cu & 5322& 1181& 14:47:43.31 & +09:39:33.9 & 119 & 300 \\
2006cy & 5159& 1426& 13:08:01.23 & +26:06:59.0 & 150 & 600 \\
2006dc & 5118& 1179& 16:16:04.04 & $-$22:37:16.7 & 107 & 300 \\
2006eh & 5140& 1127& 02:50:11.95 & +47:03:20.5 & 114 & 300 \\
2006fp & 5200& 1082& 22:45:41.13 & +73:09:47.8 & 21 & 300 \\
2006gy & 5073& 1111& 03:17:27.06 & +41:24:19.5 & 76 & 300 \\
2006jd & 5201& 1149& 08:02:07.43 & +00:48:31.5 & 77 & 300 \\
2006qq & 5108& 1048& 05:19:50.43 & $-$20:58:06.4 & 119 & 300 \\
2007K & 5081& 1069& 09:14:26.21 & +36:06:52.4 & 90 & 300 \\
2007ak & 5322& 965& 05:20:40.75 & +08:48:16.0 & 66 & 300 \\
2007bb & 5062& 953& 07:01:07.46 & +51:15:57.3 & 86 & 300 \\
2007cm & 5066& 964& 12:42:45.18 & +55:08:57.1 & 66 & 300 \\
2007pk & 5322& 660& 01:31:47.07 & +33:36:54.1 & 69 & 300 \\
2007rt & 5203& 780& 11:02:34.29 & +50:34:58.5 & 93 & 300 \\
2008B & 5066& 597& 15:02:43.65 & +23:20:07.8 & 79 & 300 \\
2008J & 5135& 593& 02:34:24.20 & $-$10:50:38.5 & 66 & 300 \\
2008aj & 5322& 812& 13:33:06.33 & +33:08:59.0 & 102 & 300 \\
2008be & 5066& 636& 14:27:48.71 & +69:41:50.8 & 123 & 300 \\
2008cg & 5135& 474& 15:54:15.15 & +10:58:25.0 & 151 & 600 \\
2008en & 5065& 386& 00:55:13.56 & +35:26:26.2 & 151 & 600 \\
2008fm & 5169& 351& 23:49:03.51 & +26:47:39.3 & 157 & 600 \\
2008gm & 5199& 301& 23:14:12.39 & $-$02:46:52.4 & 49 & 300 \\
2008ip & 5088& 490& 12:57:50.20 & +36:22:33.5 & 63 & 300
\enddata
\tablenotetext{*}{All distances are derived from the host-galaxy redshift assuming $H_0$ = 72~km s$^{-1}$~Mpc$^{-1}$.}
\end{deluxetable*}

Warm $Spitzer$/IRAC \citep{fazio04} surveyed the positions of the 69 SNe~IIn in the sample. Although the $Spitzer$~warm mission is limited to IRAC bands 1 (3.6 \micron) and 2 (4.5 \micron), ground-based observations cannot compete because they have limited sensitivity past 2 \micron~given the rapidly rising thermal background.  Furthermore, these two bands span the peak of the blackbody emission from warm dust ($500 \lesssim T_{\rm d} \lesssim 1000$~K), providing potential constraints on the dust temperature and mass (and thus the luminosity).  Assuming peak luminosities comparable to that of SN 2005ip \citep{fox10} for each SN, we developed Astronomical Observation Requests (AORs) that would yield signal-to-noise ratio (SNR) $\gtrsim$ 25 at 4.5 \micron~to ensure the most-reliable extraction of the temperature and mass from the fluxes in the two bands.  All integration times were rounded up to a multiple of 5~min to produce a simple and straightforward AOR.  Table \ref{tab2} lists the observational details.

\begin{figure}
\epsscale{0.55}
\subfigure{\label{f1a} \plotone{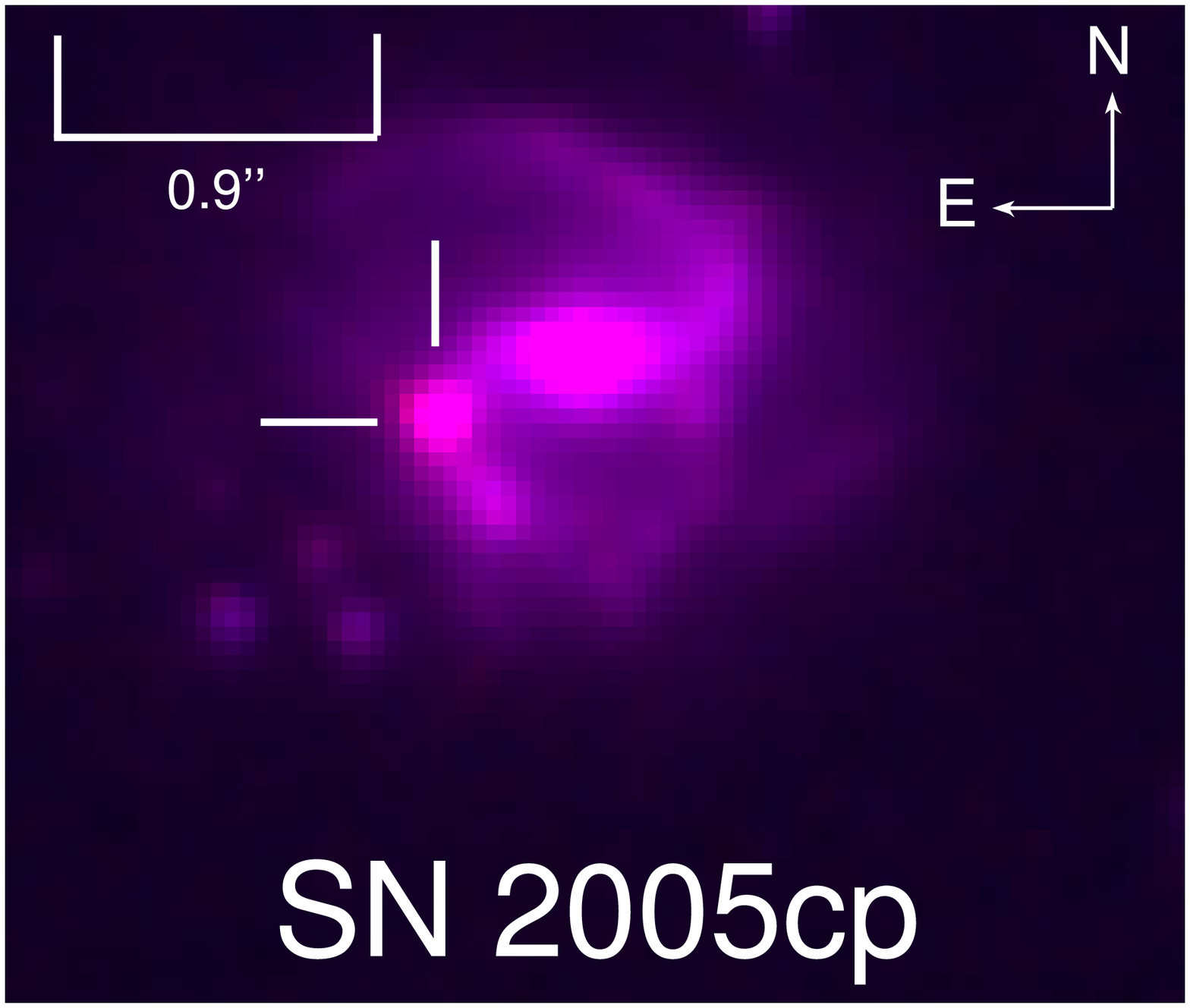}}
\subfigure{\label{f1b} \plotone{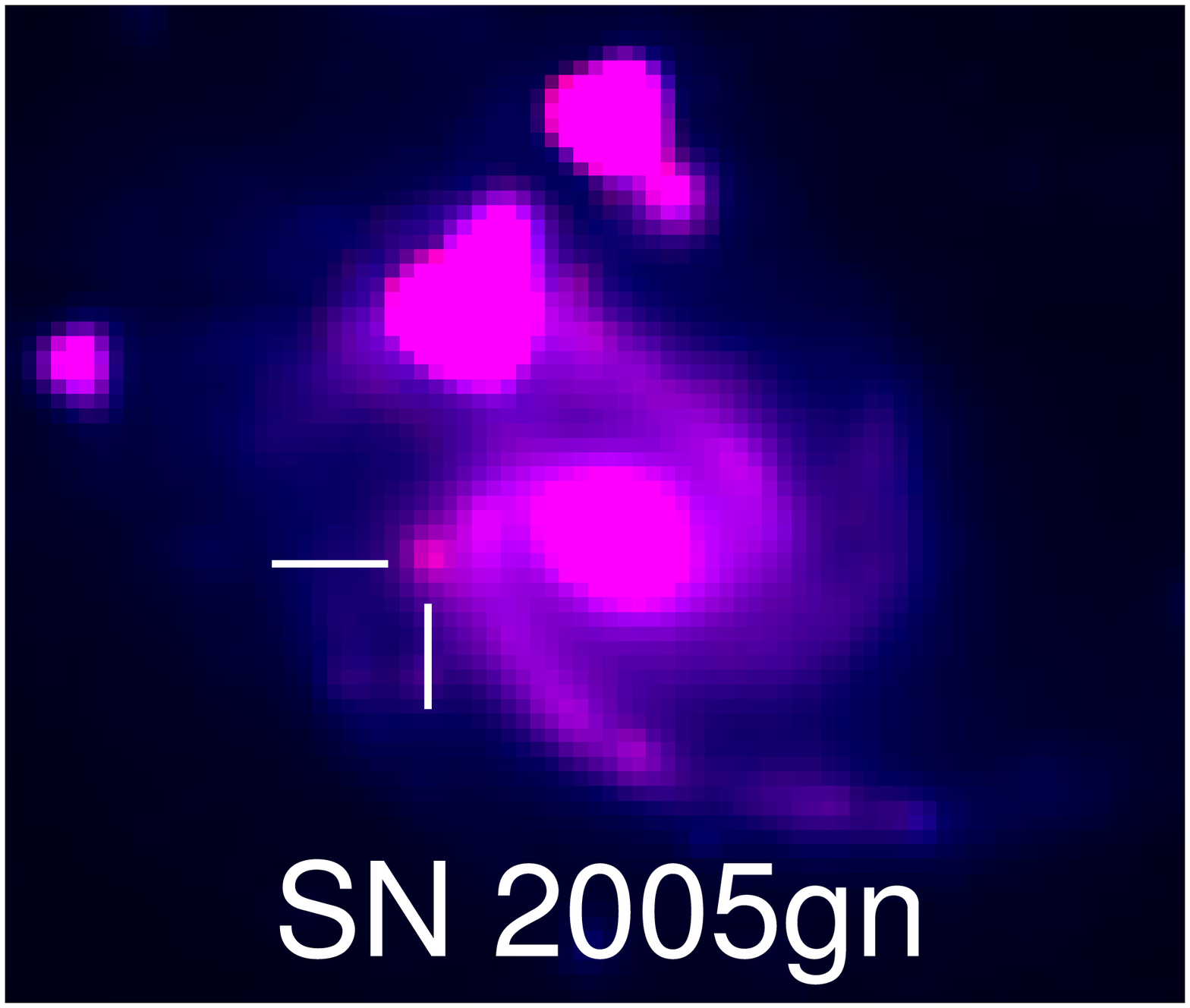}}\\
\subfigure{\label{f1c} \plotone{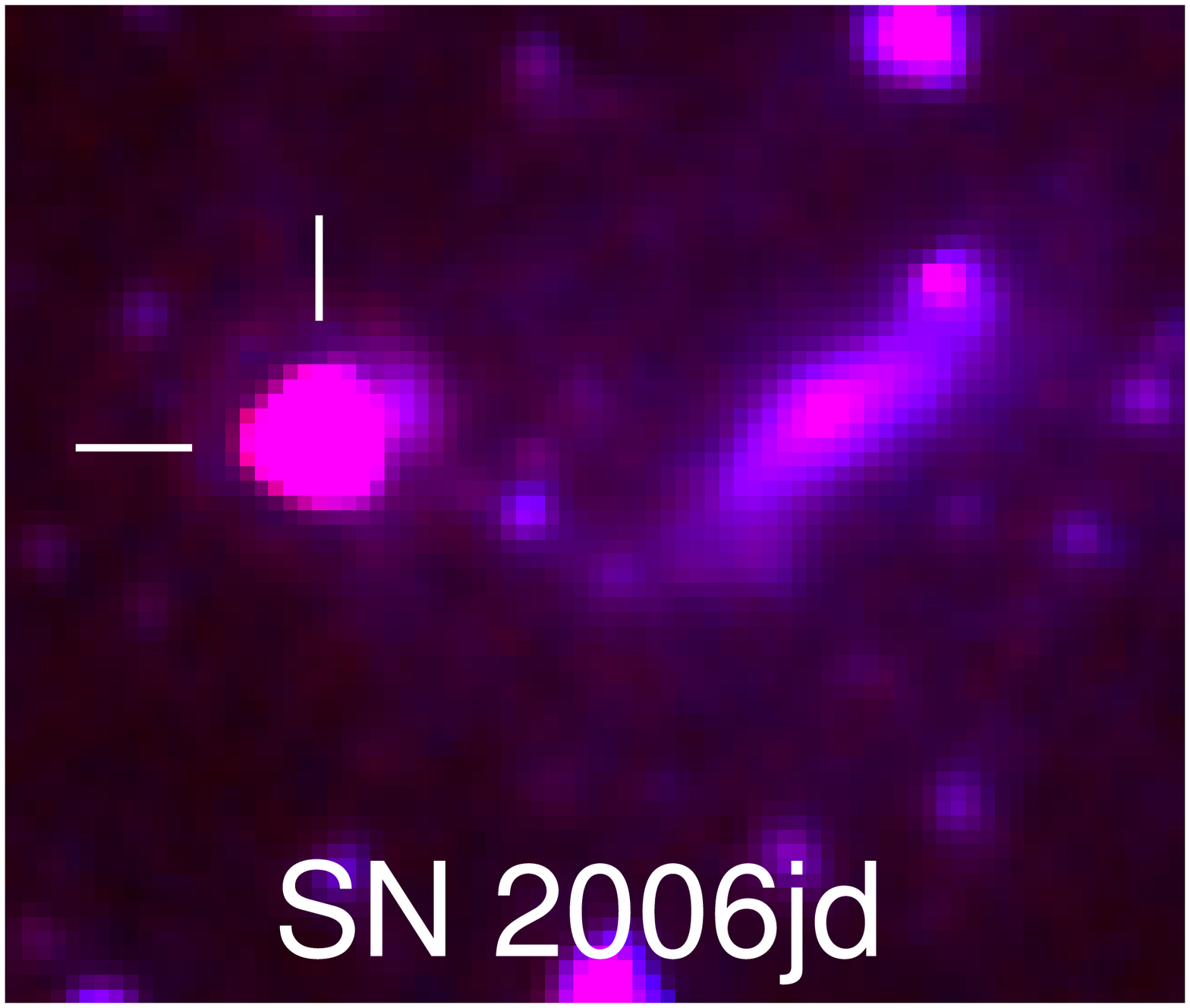}}
\subfigure{\label{f1d} \plotone{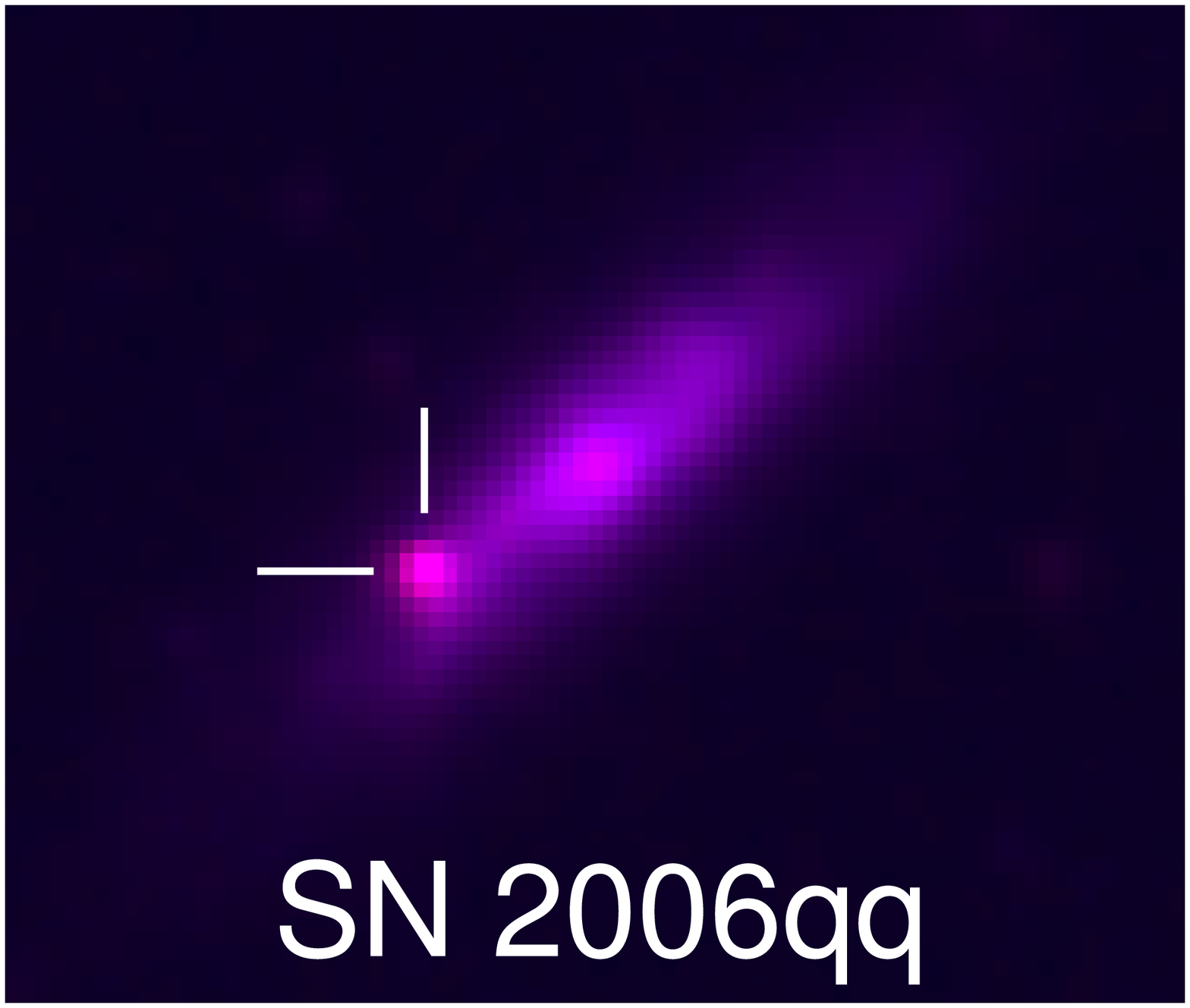}}\\
\subfigure{\label{f1e} \plotone{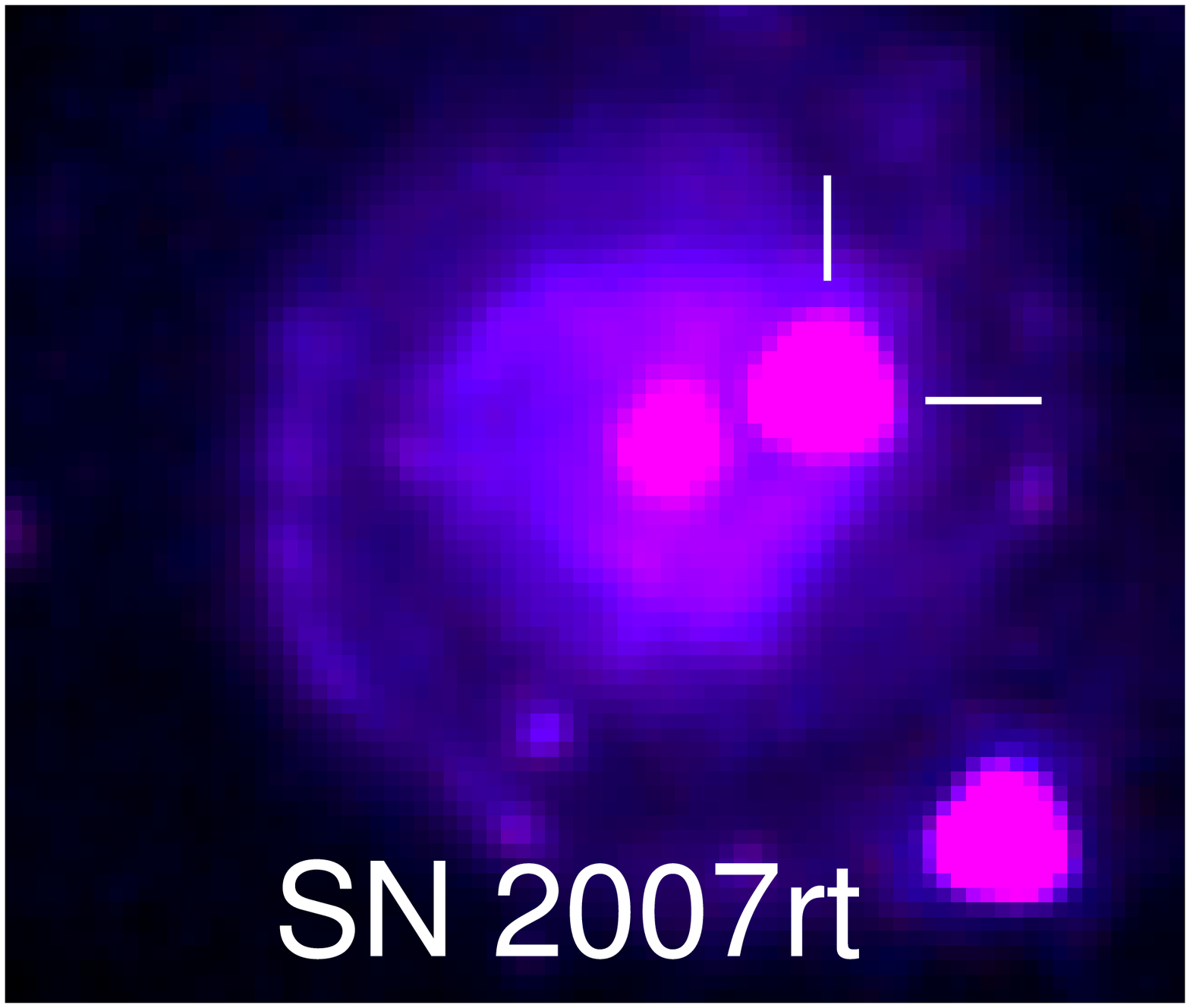}}
\subfigure{\label{f1f} \plotone{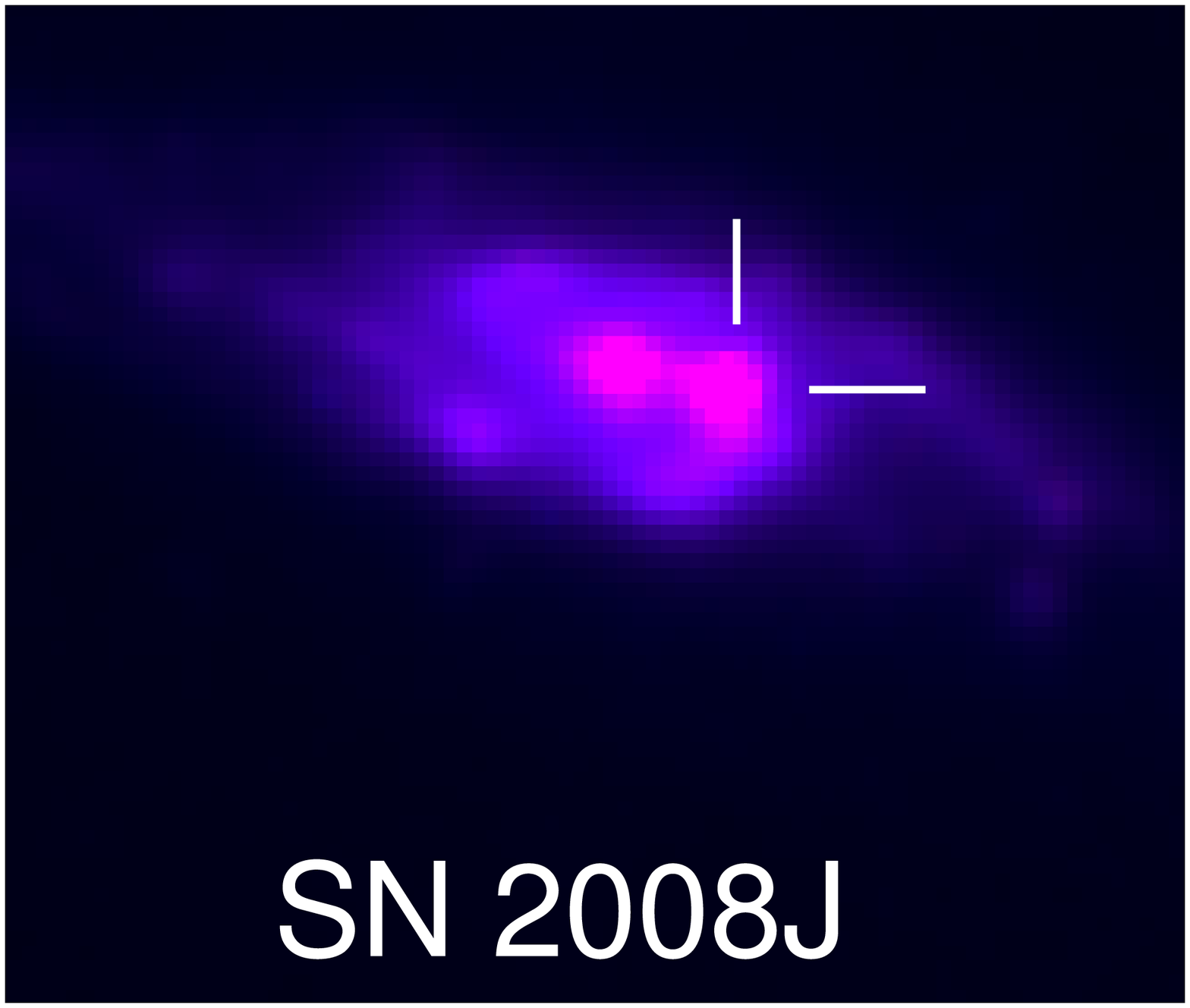}}\\
\subfigure{\label{f1g} \plotone{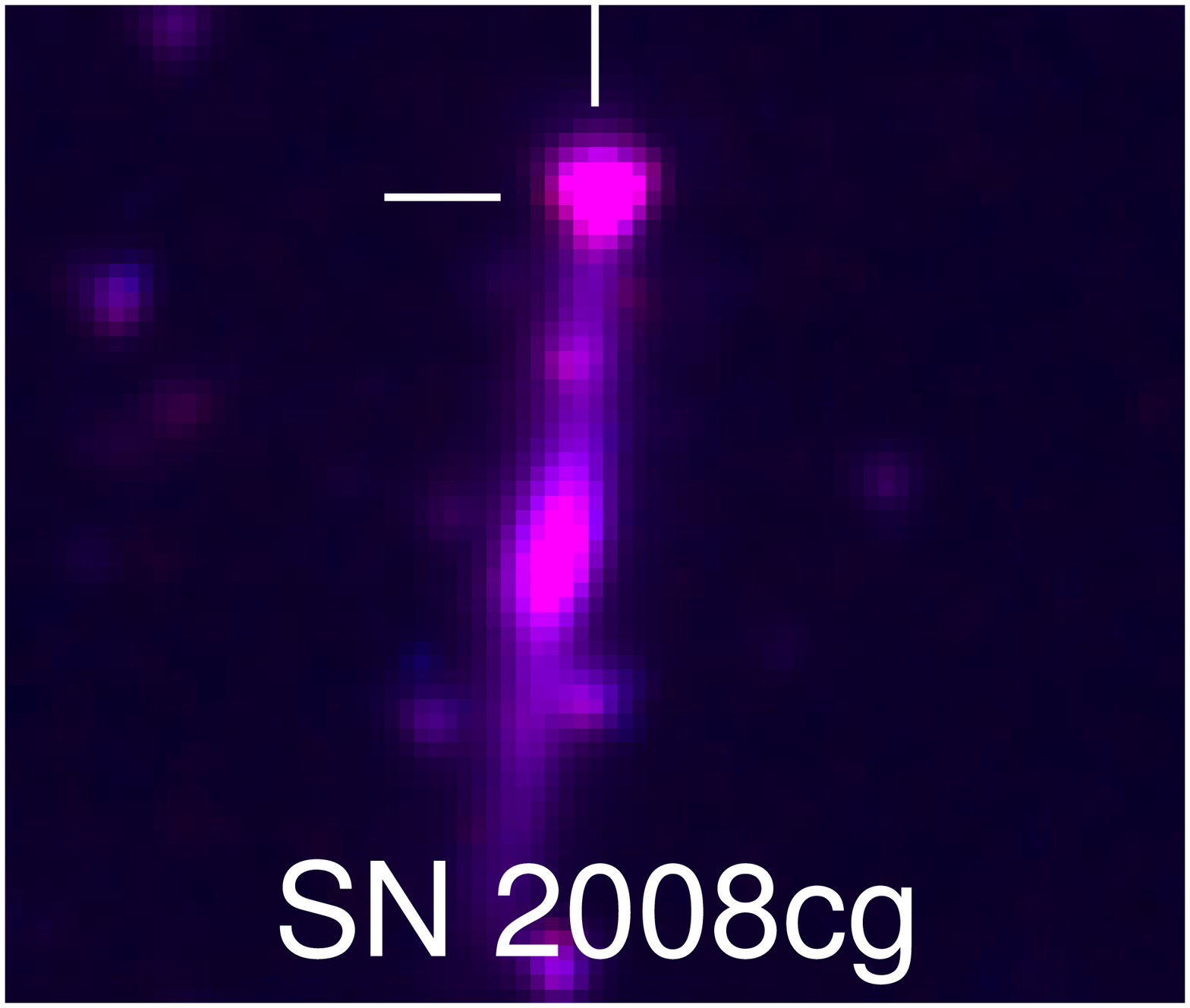}}
\subfigure{\label{f1h} \plotone{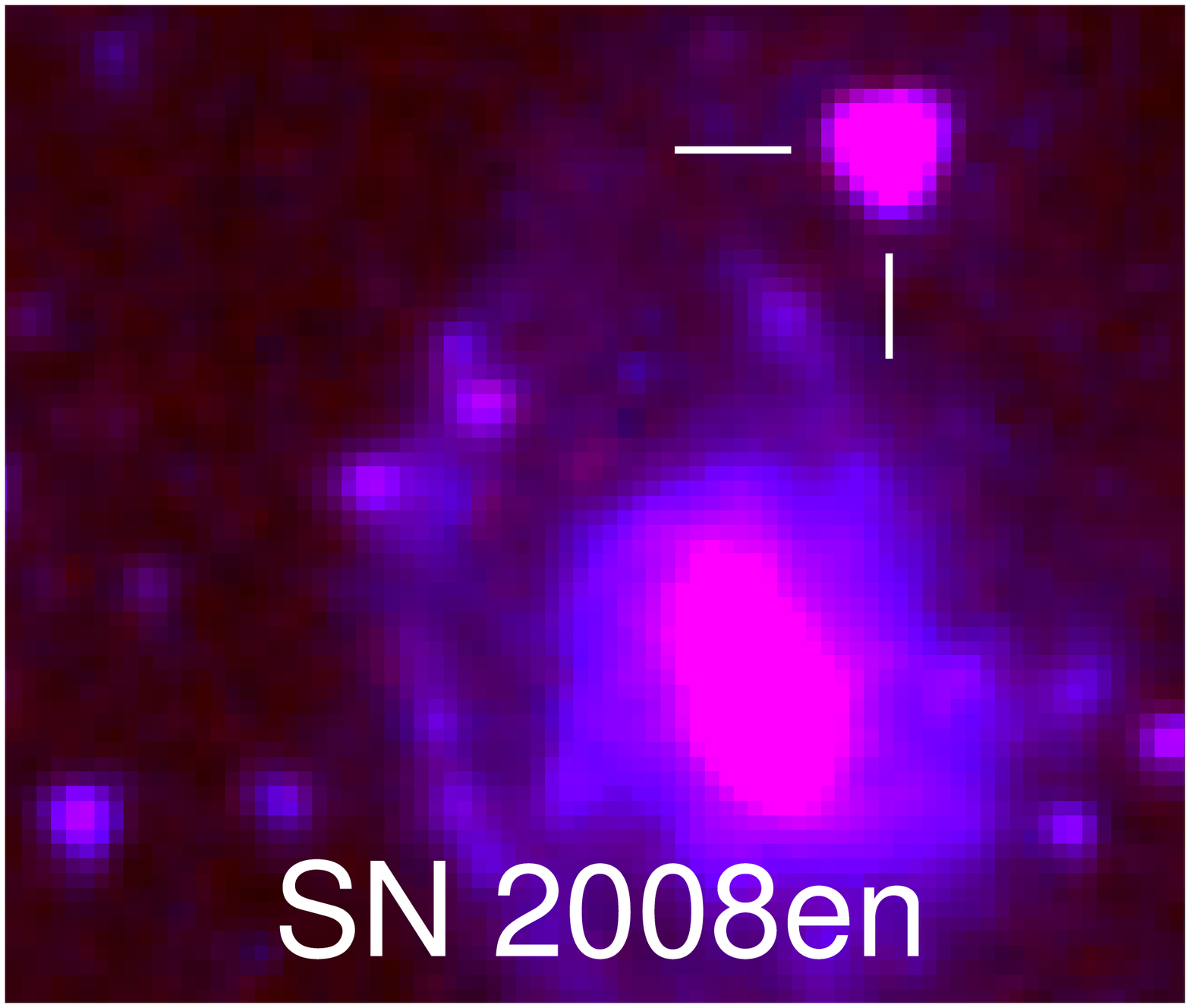}}\\
\subfigure{\label{f1i} \plotone{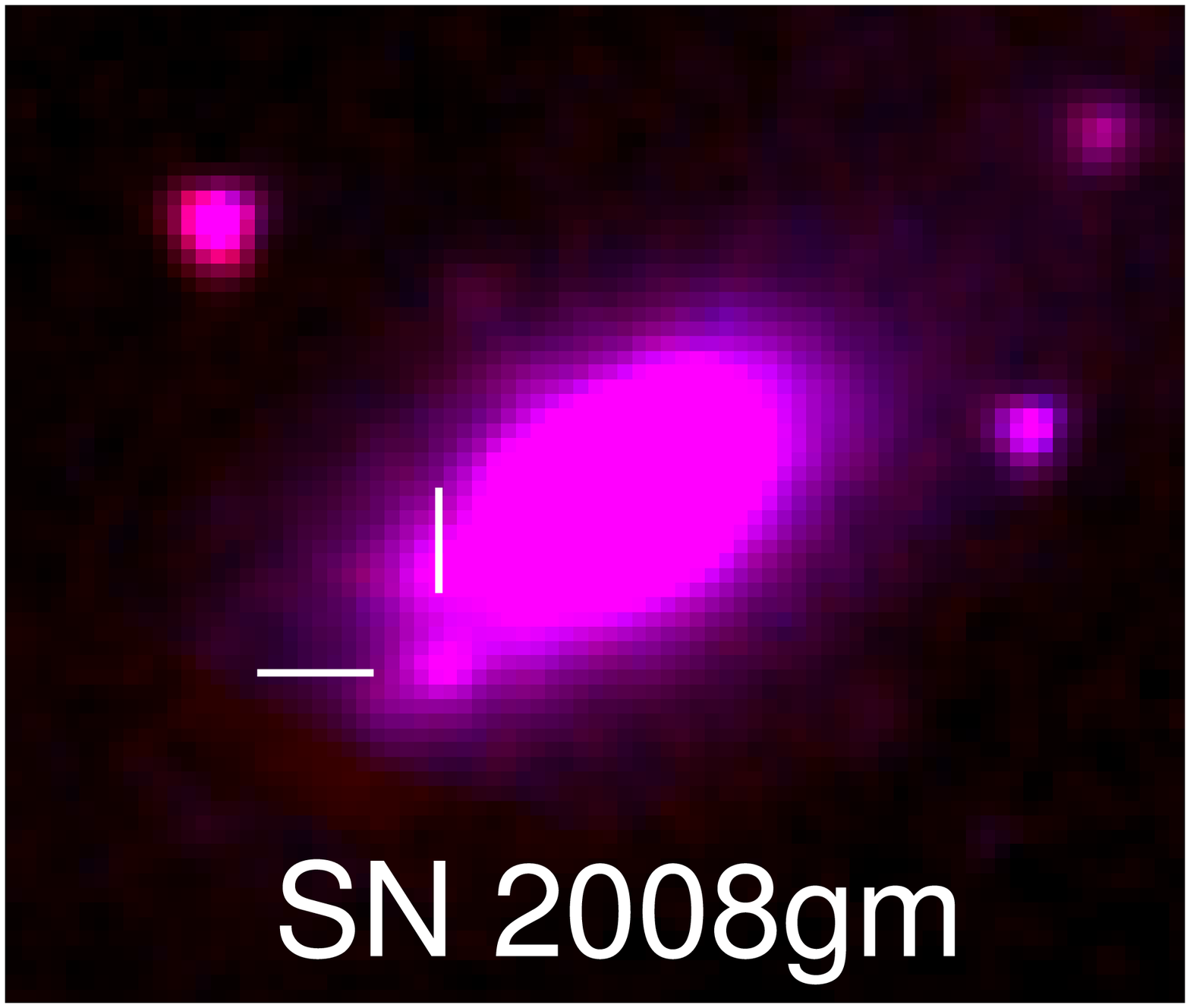}}
\subfigure{\label{f1j} \plotone{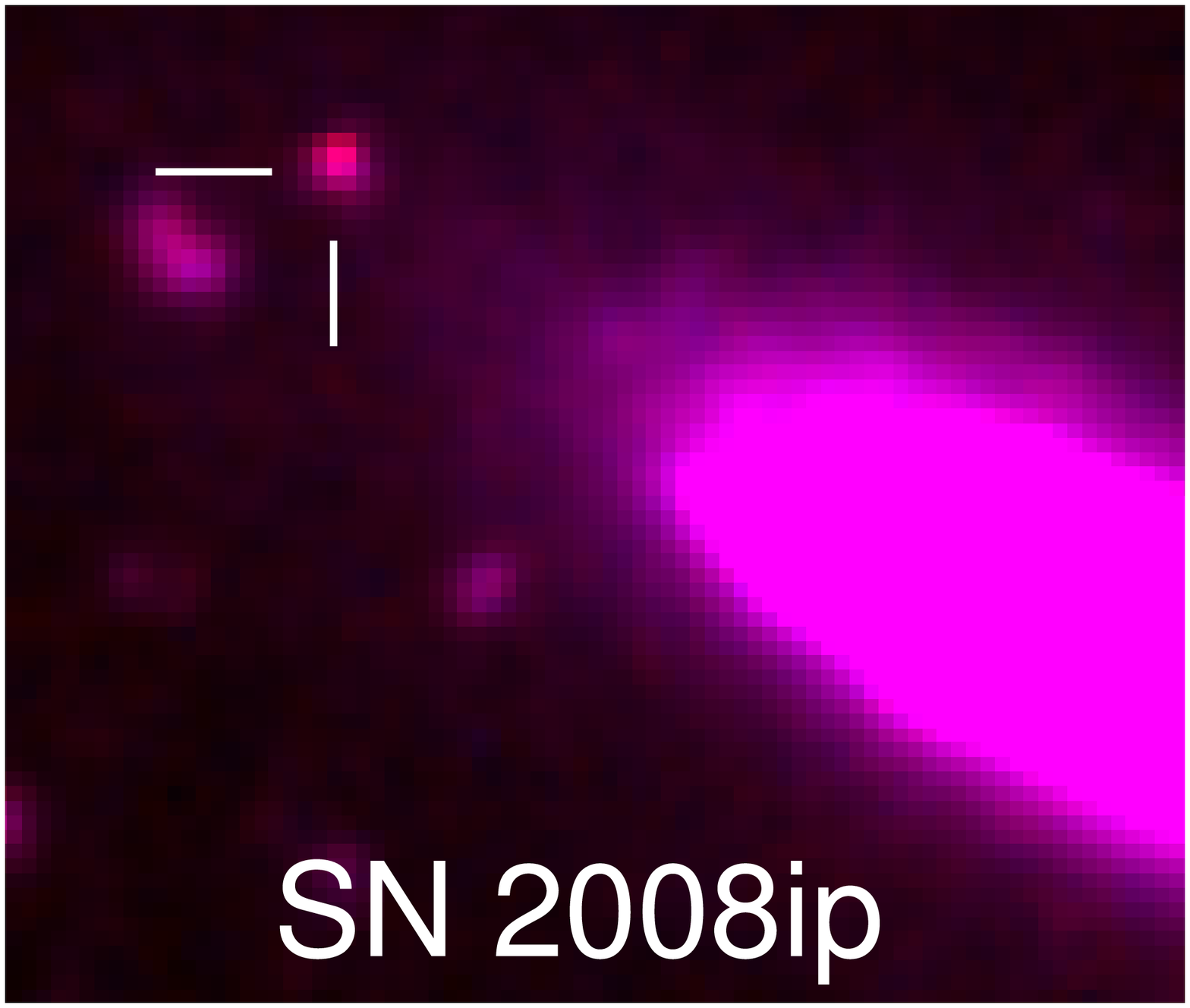}}
\caption{False-color 3.6~\micron\ and 4.5 \micron\ combined $Spitzer$/IRAC images of the 10 SNe~IIn discovered by our survey (P60122) to exhibit late-time mid-IR emission.  
}
\label{f1}
\end{figure}

The Leopard\footnote{Leopard can be downloaded from http://ssc.spitzer.caltech.edu/propkit/spot/ .} software package provided access to the Post Basic Calibrated Data ({\tt pbcd}), which are already fully coadded and calibrated.  Supernova detections were initially identified with photometry routines (SNR $\gtrsim$ 3) and later confirmed with a visual inspection.  Of the 69 SN positions surveyed, 10 SNe exhibit late-time emission (see Figure \ref{f1}).

The background flux in most of the SN host galaxies is bright and exhibits rapid spatial variations, making faint detections difficult for the {\tt DAOPHOT} point-spread function (PSF) photometry package in {\tt IRAF}\footnote{IRAF: the Image Reduction and Analysis Facility is distributed by the National Optical Astronomy Observatory, which is operated by the Association of Universities for Research in Astronomy (AURA) under cooperative agreement with the National Science Foundation (NSF).} and limiting the accuracy of sky-brightness measurements using the annuli in the IRAF {\tt APPHOT} (aperture photometry) package.  Although template subtraction is a commonly used technique to minimize photometric confusion from the underlying galaxy, no pre-SN observations exist.

\begin{figure}
\begin{center}
\epsscale{1.15}
\plotone{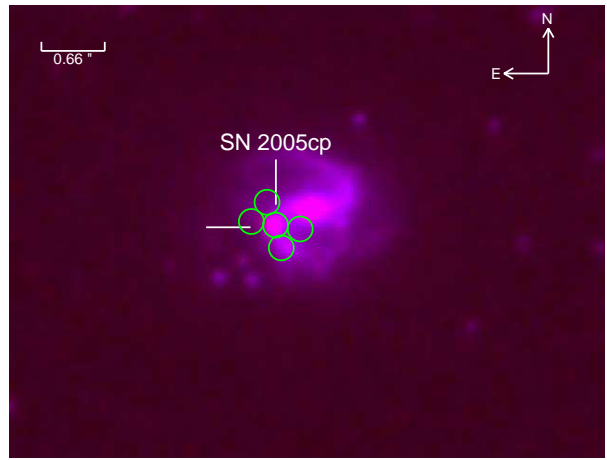}
\caption{Illustration of the aperture-photometry routine implemented instead of {\tt DAOPHOT} and {\tt APPHOT} (which utilizes both an inner aperture and outer annulus).  A set of single apertures, identified by the green circles, extracts the flux from both the SN and a user-selected background.  This technique allows the user to visually identify and extract local background representative of the underlying SN flux.  Nearby stellar and nuclear flux can be ignored.
}
\label{f2}
\end{center}
\end{figure}

Instead, photometry was performed by using a set of single apertures, with a radius defined by a fixed multiple of the PSF full width at half-maximum intensity (FWHM), to estimate both the SN and average background flux.  This technique allowed the user to visually identify only local background associated with the SN, as opposed to the annuli imposed by {\tt APPHOT}.  Figure \ref{f2} illustrates this technique in the case of SN 2005cp, where the circles identify the apertures used to extract both the SN and background fluxes.  Nearby stellar and nuclear flux, which would normally be included in the sky annuli, do not significantly contribute to the SN flux and are ignored by this technique.  

\begin{figure}
\begin{center}
\epsscale{1.2}
\plotone{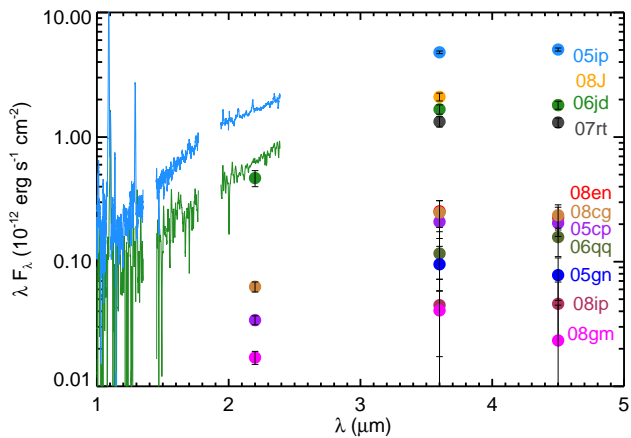}
\caption{Photometry of the SNe detected by $Spitzer$/IRAC bands 1 (3.6~\micron) and 2 (4.5 \micron), as well as near-IR photometry obtained by Magellan/PANIC.  Overplotted are the APO/TripleSpec near-IR spectra of SN 2006jd and SN 2005ip.
}
\label{f3}
\end{center}
\end{figure}

Pixel fluxes were converted from mJy sr$^{-1}$ to mJy according to the IRAC Data Handbook version 3, which discusses the pixel size and aperture correction in detail.  The {\tt DAOPHOT} and {\tt APPHOT} results for the brighter SNe agreed within the uncertainties.  Because {\tt DAOPHOT} did not succeed in detecting the fainter events, however, we only include results from {\tt APPHOT}. These fluxes are listed in Table \ref{tab3} and plotted in Figure \ref{f3}.  Upper limits for nondetections ($\sim$0.015 mJy at 4.5 \micron) were set by the point-source sensitivity in Table 2.10 of the IRAC Instrument Handbook version 2.  In some cases, coincident near-IR data from several other instruments exists (see below).

\subsection{$Spitzer$~Survey Supernova Detections}
\label{sec:detections}

{\bf SN 2005cp} was discovered in UGC 12886 on 2005 June 20.48 (UT dates are used throughout this paper; $R \approx 17.4$ mag) by \citet{lee05} and spectrally classified by \citet{modjaz05}.  The line profiles consist of both a narrow (FWHM $\approx$ 400 \kms) and intermediate (FWHM $\approx$ 4000 \kms) component.  \citet{kiewe10} provide additional early-time optical photometry and spectra, as well as an associated mass-loss rate.

{\bf SN 2005gn} was discovered in ESO 488-G30 on 2005 October 13.53 ($R \approx 18.0$ mag) by \citet{luckas05} and \citet{prasad05}, and spectrally classified by \citet{blanc05} and \citet{marsden05}.  The line profiles consist of narrow, intermediate, and broad features, although no line widths were reported.

{\bf SN 2006jd} was discovered in UGC 4179 on 2006 October 12.54 ($R \approx 17.2$ mag) by \citet{itagaki06} and \citet{prasad06jd}. It was classified as a SN~IIn, but no line widths were reported.  \citet{immler07jd} found an associated X-ray luminosity of $L_{\rm X} = 2.5 \times 10^{41}$~\ergs~and a visual magnitude $V = 19.1$ on 2007 November 16.73.  {\it Swift}/XRT measured an X-ray luminosity $L_{\rm X} = 2.5\times10^{41}$~\ergs~on day 400 \citep{immler07jd}, and \citet{chandra07jd} reported a radio flux density of $238~\mu$Jy on day 405.

\begin{deluxetable}{ l c c c }
\tablewidth{0pt}
\tabletypesize{\footnotesize}
\tablecaption{$Spitzer$~Survey Detections \tablenotemark{1} \label{tab3}}
\tablecolumns{4}
\tablehead{
\colhead{SN} & \colhead{Epoch} & \colhead{3.6~\micron \tablenotemark{*}} & \colhead{4.5 \micron \tablenotemark{*}}\\
\colhead{ } & \colhead{(days)} & \colhead{(mJy)} & \colhead{(mJy)}
}
\startdata
2005cp & 1523 & 0.25(0.07) & 0.31(0.07) \\
2005gn & 1479 & 0.11(0.04) & 0.12(0.04) \\
2005ip & 948 & 5.76(0.14) & 7.54(0.22) \\
2006jd & 1149 & 2.00(0.17) & 2.71(0.20)\\
2006qq & 1048 & 0.14(0.07) & 0.24(0.07) \\
2007rt & 780 & 1.60(0.15) & 1.96(0.17) \\
2008J & 593 & 2.52(0.18) & 2.71(0.19) \\
2008cg & 474 & 0.30(0.07) & 0.35(0.08) \\
2008en & 386 & 0.30(0.07) & 0.34(0.07) \\
2008gm & 301 & 0.05(0.04) & 0.03(0.03) \\
2008ip & 438 & 0.05(0.03) & 0.07(0.03) \\
\enddata
\tablenotetext{*}{1$\sigma$ uncertainties are given in parentheses.}
\tablenotetext{1}{Upper limits for nondetections ($\sim$0.015 mJy at 4.5 \micron) were set by the point-source sensitivity in Table 2.10 of the IRAC Instrument Handbook version 2.}
\end{deluxetable}

{\bf SN 2006qq} was discovered in ESO 553-G36 on 2006 October 27.40 ($R \approx 17.0$, $M_V = -18.4$ mag) by \citet{prasad06qq} and classified by \citet{silverman06}.  The line profiles consist of narrow (FWHM $\approx$ 900 \kms) emission upon a relatively featureless continuum.

{\bf SN 2007rt} was discovered in UGC 6109 on 2007 November 24.54 ($R \approx 16.8$ mag) by \citet{li07}. It was spectrally classified as a SN~IIn about 2--3 months past maximum light \citep{blondin07}.  The line profiles consist of both an intermediate (FWHM $\approx$ 2500 \kms) and broad (FWHM $\approx$ 10,000 \kms) component.  A moderate-resolution spectrum reveals an additional narrow H$ \alpha$~P-Cygni profile (FWHM $\approx$ 100 \kms) and evidence for new dust formation \citep{trundle09}.  On day 48, \citet{chandra08rt} did not detect any radio emission and assigned an upper limit of $< 9~\mu$Jy.

{\bf SN 2008J} was discovered in MCG--02-7-33 on 2008 February 15.19 ($R \approx 15.4$ mag) by \citet{thrasher08} and classified by \citep{green08J}.  The line profiles consist of both a narrow and intermediate component, although the line widths were not listed.

{\bf SN 2008cg} was discovered in FGC 1965 on 2008 May 5.39 ($R \approx 16.7$) by \citet{drake08} and classified by \citep{blondin08}.  The line profiles consist of both a narrow (FWHM $\approx$ 300 \kms) and intermediate (FWHM $\approx$ 1500 \kms) component.  {\it Swift}/UVOT measured $V = 17.9$ mag on day 58 \citep{immler08cg}.  On day 57, however, \citet{chandra08cg} did not detect any radio emission and assigned an upper limit of $< 7~\mu$Jy.

{\bf SN 2008en} was discovered in UGC 564 on 2008 August 3.06 ($R \approx 18.2$ mag) by \citet{boles08} and spectrally classified by \citet{green08en}.  These spectra reveal broad (FWHM $\approx$ 3700 \kms) Balmer and He~I emission lines.

{\bf SN 2008gm} was discovered in NGC 7530 on 2008 October 22.09 ($R \approx 17.0$ mag) by \citet{pignata08} and classified by \citet{prieto08}.  The line profiles consist of both a narrow (FWHM $\approx$ 300 \kms) and intermediate (FWHM $\approx$ 1600 \kms) component. Two days following the discovery, \citet{soderberg08gm} did not detect any radio emission and assigned an upper limit of $<$ 0.08 mJy (3$\sigma$).

{\bf SN 2008ip} was discovered in NGC 4846 on 2008 December 31.75 ($R \approx 15.7$ mag) by \citet{nakano09ip} and classified by \citet{challis09ip}.  No spectral line widths were given.  \citet{chandra09ip} report no radio counterpart on day 5 and assign a 3$\sigma$~upper limit of 0.096 mJy (8.46 GHz).

\subsection{Near-Infrared Photometry and Spectroscopy}
\label{sec:nir}

\begin{deluxetable}{ l c c c c c }
\tablewidth{0pt}
\tabletypesize{\footnotesize}
\tablecaption{Summary of Optical Spectra \label{tab4}}
\tablecolumns{6}
\tablehead{
\multirow{2}*{SN} & Epoch & \multirow{2}*{Instrument} & Blue & Red & Int\\
& (days) & & Resolution & Resolution& (s)
}
\startdata
2005cp & 10 & Lick/Kast & 7.7 & 10.4 & 900 \\
\hline
\multirow{2}*{2006jd} & 395 & Keck/LRIS & 6.5 & 7 & 900 \\
	& 564 & Keck/LRIS & 6.5 & 5.8 & 600 \\
\hline
\multirow{2}*{2006qq}& 1 & Lick/Kast & 6 & 10.9 & 1800 \\
& 20 & Lick/Kast & 6.7 & 10.0 & 2100 \\
\hline
\multirow{7}*{2008J} & 32 & Lick/Kast & 4.9 & 11.9 & 900 \\
& 199 & Lick/Kast & 6 & 11 & 900 \\
& 224 & Lick/Kast & 6.7 & 12.3 & 900 \\
& 236 & Lick/Kast & 7.4 & 12.3 & 1200 \\
& 251 & Lick/Kast & 4.6 & 11.0 & 1200 \\
& 266 & Lick/Kast & 4.4 & 11.0 & 1200 \\
& 313 & Lick/Kast & 4.8 & 5 & 1200 \\
\hline
\multirow{5}*{2008cg}& 3 & Lick/Kast & 5.3 & 11.6 &1800 \\
& 10 & Lick/Kast & 5.5 & 11.7 & 1800\\
& 55 & Lick/Kast & 5.9 & 10.8 & 1500 \\
& 63 & Lick/Kast & 6 & 11.3 & 1500\\
& 114 & Keck/LRIS & 4.5 & 7 & 454 \\
\hline
2008en & 5 & Lick/Kast & 6.7 & 12.5 & 1800 \\
\hline
2008gm & 5 & Keck/LRIS & 6.5 & 7 & 400
\enddata
\end{deluxetable}

Nearly coincident near-IR photometric data were collected for several targets with the Persson Auxiliary Nasmyth Infrared Camera \citep[PANIC;][]{martini04} mounted on the 6.5-m Magellan (Walter Baade) Telescope at the Las Campanas Observatory, Chile.  PANIC obtained $J$- and $K_s$-band photometry of SNe 2008gm (363 days post-discovery), 2008cg (659 days), 2006jd (1230 days), and 2005cp (1582 days).  Standard near-IR data-reduction techniques were implemented.  The relatively small galaxy sizes allow data frames to double as sky exposures, provided a sufficiently large dither avoids an overlapping galaxy from one exposure to the next.  Photometry was performed with the DAOPHOT PSF-fitting techniques. The results are plotted in Figure \ref{f3} adjacent to the $Spitzer$~photometry and, together, are treated as a single epoch given the near simultaneity of the observations relative to the overall epoch of observation.

TripleSpec, a 0.9--2.5 \micron, resolution $R=3000$ spectrograph operating at Apache Point Observatory in NM \citep{wilson04,herter08}, was used to obtain a spectrum of SN 2006jd on day 1250 post-discovery and a spectrum of SN 2005ip on day 862 post-discovery (originally published by \citealt{fox10}).  Forty minutes of on-source integration consisted of 8 independent 5~min exposures nodding between 2 different slit positions.  We extracted the spectra with a modified version of the IDL-based {\it SpexTool} \citep{cushing04}.  Any underlying galactic and sky emission are approximated in {\it SpexTool} by a polynomial fit and subtracted from the SN.  The results are again plotted in Figure \ref{f3} and, combined with the $Spitzer$~data, treated as a single epoch.

\subsection{Optical Photometry and Spectroscopy}
\label{sec:opt}

Table \ref{tab4} summarizes the optical spectra obtained with various instruments.  Data were obtained with the dual-arm Low Resolution Imaging Spectrometer \citep[LRIS;][]{oke95} mounted on the 10-m Keck~I telescope, as well as with the Kast double spectrograph \citep{miller93} mounted on the Shane 3-m telescope at Lick Observatory.  Keck/LRIS spectra were obtained using the 600/4000 or 400/3400 grisms on the blue side and the 400/8500 grating on the red side, along with a 1\arcsec~wide slit. This resulted in a wavelength coverage of 3200--9200~\AA\ and a typical resolution of 5--7~\AA. Almost every Lick/Kast spectrum was obtained using the 600/4310 grism on the blue side and the 300/7500 grating on the red side, along with a 2\arcsec~wide slit.  This resulted in a wavelength coverage of 3300--10,400~\AA\ and resolutions of about 5--7~\AA\ and 10--12~\AA\ on the blue and red sides, respectively. Most observations had the slit aligned along the parallactic angle to minimize differential light losses \citep{filippenko82}.

The spectra were reduced using standard techniques \citep[e.g.,][]{foley03}. Routine CCD processing and spectrum extraction were completed with IRAF, and the data were extracted with the optimal algorithm of \citet{horne86}. We obtained the wavelength scale from low-order polynomial fits to calibration-lamp spectra. Small wavelength shifts were then applied to the data after cross-correlating a template sky to an extracted night-sky spectrum. Using our own IDL routines, we fit a spectrophotometric standard-star spectrum to the data in order to flux calibrate the SN and to remove telluric absorption lines \citep{wade88,matheson00}.  

\begin{figure}
\begin{center}
\epsscale{1}
\plotone{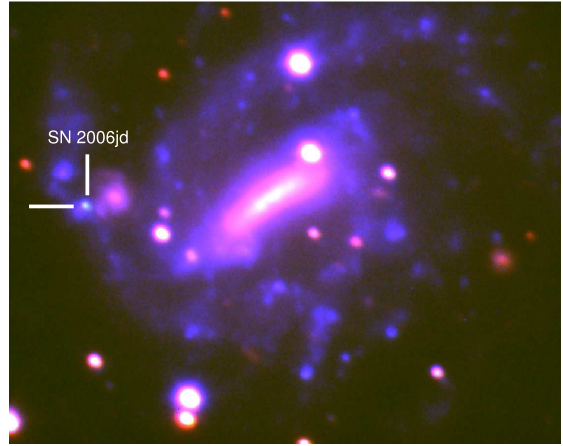}
\caption{False-color $BRI$~combined Keck/LRIS images of SN 2006jd on day 1609 post-detection.  The SN remains bright at optical wavelengths nearly 4.5 yr after the explosion.  
}
\label{f4}
\end{center}
\end{figure}

Keck/LRIS also obtained $BVRI$ photometry of SNe 2006jd (1609 days) and 2007rt (1223 days) on 2011 March 9.  Observations of each target consisted of two dithered 120-s exposures in each band.  Again, the data were reduced using standard techniques and routine CCD processing with IRAF.  Table \ref{tab5} lists the photometric results.  Only SN 2006jd was detected (see Figure \ref{f4}), but upper limits for SN 2007rt are also provided.  

\begin{deluxetable}{ l c c c c c}
\tablewidth{0pt}
\tabletypesize{\footnotesize}
\tablecaption{Summary of Keck/LRIS Optical Photometry \label{tab5}}
\tablecolumns{6}
\tablehead{
\multirow{2}*{SN} & Epoch & $B$ & $R$ & $I$ & Luminosity \\
& (days) & (mag) & (mag) & (mag) & (log ($L_{\rm opt}$/\lsolar))
}
\startdata 
2006jd & 1609 & 25.0$\pm0.5$ & 20.3$\pm$0.5 & 22.5$\pm0.5$ & 7.6$\pm0.4$\\
2007rt & 1223 & $>$25 & $>$24 & $>$24 & $<$6.1
\enddata
\end{deluxetable}

\subsection{Dust Composition, Temperature, and Mass}
\label{sec:dust}

\begin{figure*}
\begin{center}
\epsscale{0.48}
\subfigure[Graphite]{\label{f5a} \plotone{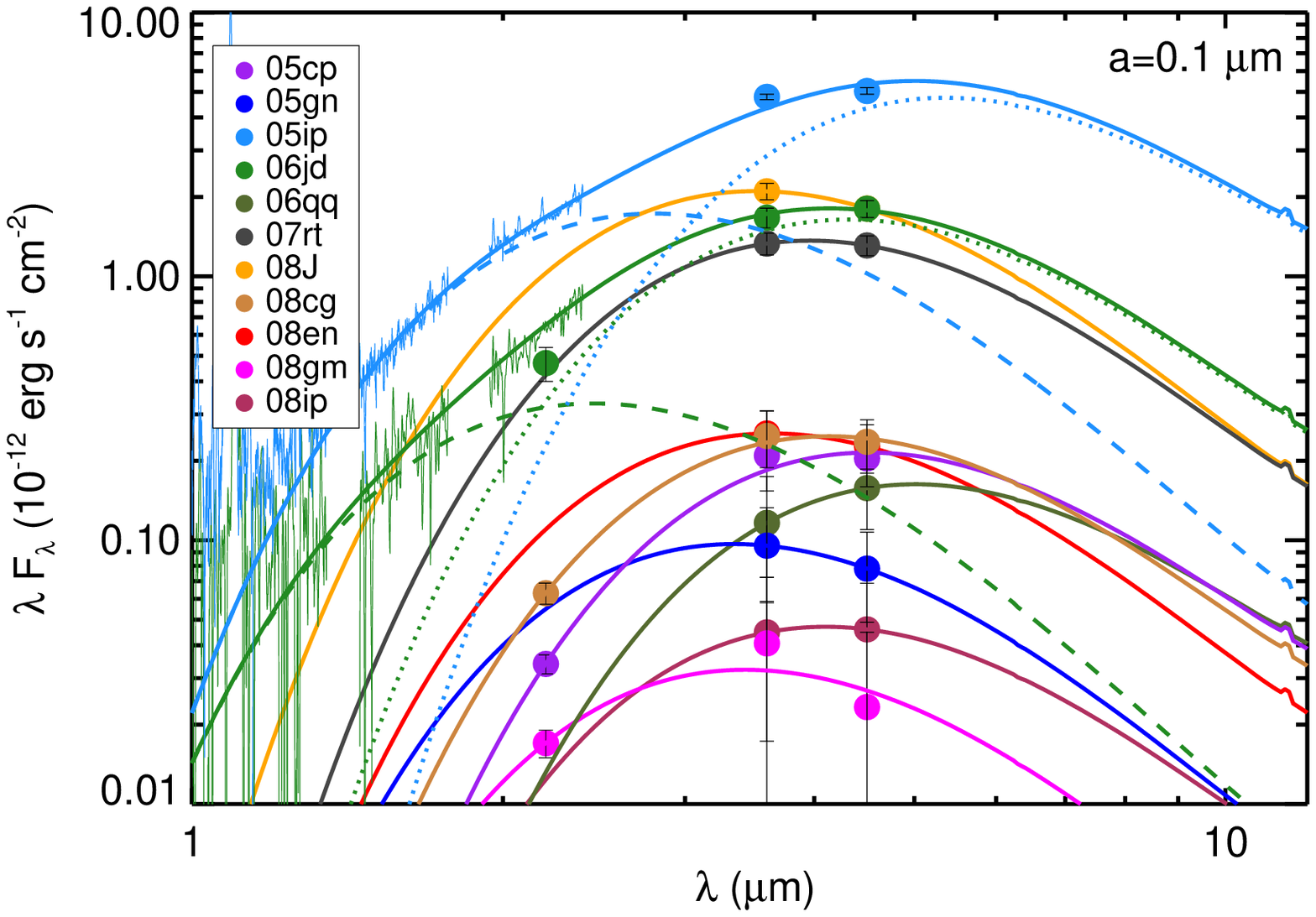}}
\subfigure[Silicate]{\label{f5b} \plotone{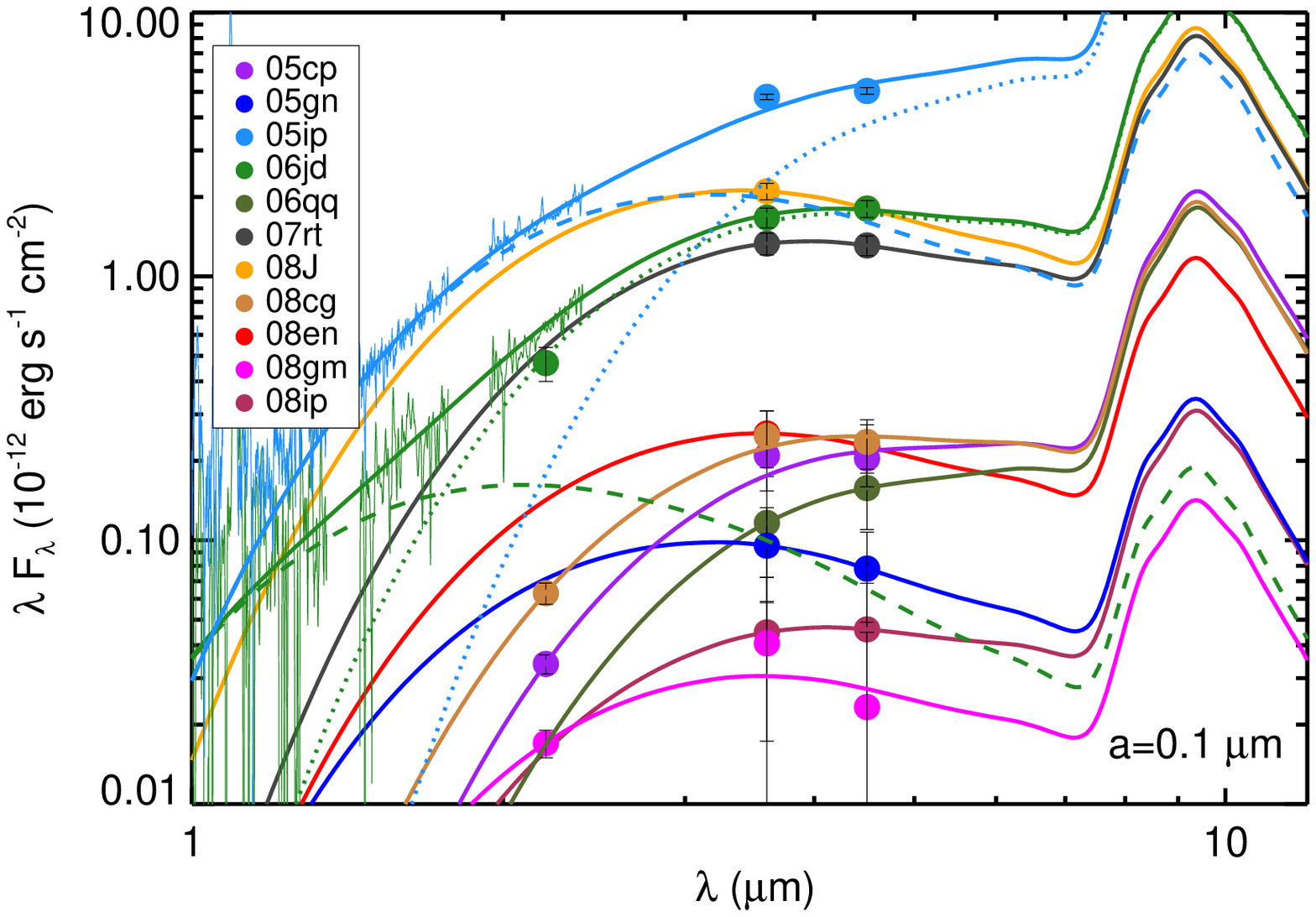}}
\caption{Best fits of Equation \ref{eqn4:flux2} (both graphite and silicate compositions) to the combined near-IR and mid-IR data.  A warm ($\sim 400$--800~K) dust component fits most observations well.  SNe 2006jd and 2005ip (those with near-IR spectra) are better fit by a multi-component model, where a second, hotter component dominates the near-IR flux.  The most obvious feature differentiating between the graphite and silicate models is the 9~\micron\ silicate emission feature.  The warm IRAC data do not extend beyond $\sim 5$ \micron, limiting any distinction between the two compositions.
}
\label{f5}
\end{center}
\end{figure*}

\begin{deluxetable*}{c c c c c c c c c c}
\tablewidth{0pt}
\tabletypesize{\footnotesize}
\tablecaption{$Spitzer$~Fitting Parameters \label{tab6}}
\tablecolumns{10}
\tablehead{ 
\colhead{} & \multicolumn{3}{c}{Graphites} & \multicolumn{3}{c}{Silicates}& \multicolumn{3}{c}{Blackbody}\\
\colhead{$a$ (\micron)} & \colhead{$T_{\rm d}$ (K)} & \colhead{$M_{\rm d}$ (\msolar)} & \colhead{$L_{\rm d}$ (\lsolar)} & \colhead{$T_{\rm d}$ (K)} & \colhead{$M_{\rm d}$ (\msolar)} & \colhead{$L_{\rm d}$ (\lsolar)} & \colhead{$T_{\rm d}$ (K)} & \colhead{$M_{\rm d}$ (\msolar)} & \colhead{$L_{\rm d}$ (\lsolar)}
}
\startdata
\multicolumn{10}{c}{\textbf{SN 2005cp}}\\
\hline
0.01 & 575 & 7.39e-3 & 5.65e+7 & 648 & 1.51e-2 & 2.06e+8 & 710 & 9.09e-6 & 6.86e+7 \\
0.10 & 554 & 6.81e-3 & 5.63e+7 & 643 & 1.55e-2 & 2.07e+8 & 710 & 9.08e-5 & 6.85e+7 \\
0.30 & 580 & 1.79e-3 & 5.15e+7 & 614 & 1.78e-2 & 2.13e+8 & 710 & 2.70e-4 & 6.79e+7 \\
\hline
\multicolumn{10}{c}{\textbf{SN 2005gn}}\\
\hline
0.01 & 776 & 2.06e-3 & 8.39e+7 & 991 & 3.12e-3 & 1.72e+8 & 1220 & 1.52e-6 & 1.06e+8 \\
0.10 & 737 & 1.92e-3 & 8.36e+7 & 981 & 3.18e-3 & 1.73e+8 & 1220 & 1.51e-5 & 1.06e+8 \\
0.30 & 735 & 6.75e-4 & 7.52e+7 & 917 & 3.67e-3 & 1.75e+8 & 1220 & 4.50e-5 & 1.05e+8 \\
\hline
\multicolumn{10}{c}{\textbf{SN 2005ip}}\\
\hline
0.01 & 483 & 4.73e-2 & 1.36e+8 & 525 & 9.12e-2 & 6.23e+8 & 567 & 5.12e-5 & 1.49e+8 \\
0.10 & 468 & 4.34e-2 & 1.35e+8 & 522 & 9.30e-2 & 6.26e+8 & 567 & 5.11e-4 & 1.48e+8 \\
0.30 & 469 & 1.46e-2 & 1.19e+8 & 503 & 1.07e-1 & 6.50e+8 & 567 & 1.52e-3 & 1.47e+8 \\
\hline
\multicolumn{10}{c}{\textbf{SN 2006jd}}\\
\hline
0.01 & 602 & 3.21e-2 & 3.18e+8 & 733 & 4.90e-2 & 9.94e+8 & 836 & 2.70e-5 & 4.03e+8 \\
0.10 & 579 & 2.96e-2 & 3.17e+8 & 728 & 5.00e-2 & 1.00e+9 & 836 & 2.70e-4 & 4.02e+8 \\
0.30 & 561 & 1.13e-2 & 2.65e+8 & 692 & 5.76e-2 & 1.02e+9 & 836 & 8.01e-4 & 3.98e+8 \\
\hline
\multicolumn{10}{c}{\textbf{SN 2006qq}}\\
\hline
0.01 & 518 & 1.71e-2 & 7.30e+7 & 601 & 2.75e-2 & 2.94e+8 & 672 & 1.45e-5 & 8.68e+7 \\
0.10 & 501 & 1.58e-2 & 7.27e+7 & 598 & 2.79e-2 & 2.95e+8 & 672 & 1.45e-4 & 8.66e+7 \\
0.30 & 500 & 5.54e-3 & 6.66e+7 & 575 & 3.16e-2 & 3.03e+8 & 672 & 4.30e-4 & 8.58e+7 \\
\hline
\multicolumn{10}{c}{\textbf{SN 2007rt}}\\
\hline
0.01 & 653 & 2.39e-2 & 3.71e+8 & 793 & 3.76e-2 & 9.84e+8 & 926 & 1.92e-5 & 4.37e+8 \\
0.10 & 625 & 2.22e-2 & 3.70e+8 & 787 & 3.83e-2 & 9.87e+8 & 926 & 1.92e-4 & 4.36e+8 \\
0.30 & 624 & 7.79e-3 & 3.41e+8 & 747 & 4.37e-2 & 1.00e+9 & 926 & 5.70e-4 & 4.32e+8 \\
\hline
\multicolumn{10}{c}{\textbf{SN 2008J}}\\
\hline
0.01 & 739 & 9.41e-3 & 2.92e+8 & 930 & 1.44e-2 & 6.37e+8 & 1125 & 7.13e-6 & 3.58e+8 \\
0.10 & 704 & 8.73e-3 & 2.91e+8 & 921 & 1.47e-2 & 6.40e+8 & 1125 & 7.12e-5 & 3.58e+8 \\
0.30 & 702 & 3.07e-3 & 2.64e+8 & 865 & 1.69e-2 & 6.48e+8 & 1125 & 2.12e-4 & 3.54e+8 \\
\hline
\multicolumn{10}{c}{\textbf{SN 2008cg}}\\
\hline
0.01 & 629 & 1.43e-2 & 1.80e+8 & 716 & 2.93e-2 & 5.52e+8 & 975 & 7.70e-6 & 2.17e+8 \\
0.10 & 604 & 1.32e-2 & 1.79e+8 & 710 & 3.02e-2 & 5.57e+8 & 975 & 7.69e-5 & 2.16e+8 \\
0.30 & 635 & 3.43e-3 & 1.67e+8 & 675 & 3.48e-2 & 5.70e+8 & 975 & 2.28e-4 & 2.14e+8 \\
\hline
\multicolumn{10}{c}{\textbf{SN 2008en}}\\
\hline
0.01 & 714 & 7.23e-3 & 1.84e+8 & 888 & 1.11e-2 & 4.23e+8 & 1062 & 5.58e-6 & 2.23e+8 \\
0.10 & 681 & 6.70e-3 & 1.84e+8 & 881 & 1.14e-2 & 4.25e+8 & 1062 & 5.57e-5 & 2.22e+8 \\
0.30 & 679 & 2.36e-3 & 1.68e+8 & 829 & 1.30e-2 & 4.31e+8 & 1062 & 1.66e-4 & 2.20e+8 \\
\hline
\multicolumn{10}{c}{\textbf{SN 2008gm}}\\
\hline
0.01 & 757 & 7.00e-5 & 2.47e+6 & 890 & 1.40e-4 & 5.36e+6 & 1013 & 8.31e-8 & 2.73e+6 \\
0.10 & 721 & 6.47e-5 & 2.46e+6 & 880 & 1.45e-4 & 5.39e+6 & 1013 & 8.29e-7 & 2.73e+6 \\
0.30 & 766 & 1.69e-5 & 2.36e+6 & 827 & 1.68e-4 & 5.48e+6 & 1013 & 2.46e-6 & 2.70e+6 \\
\hline
\multicolumn{10}{c}{\textbf{SN 2008ip}}\\
\hline
0.01 & 634 & 4.52e-4 & 5.93e+6 & 765 & 7.12e-4 & 1.66e+7 & 886 & 3.66e-7 & 6.96e+6 \\
0.10 & 608 & 4.18e-4 & 5.91e+6 & 759 & 7.25e-4 & 1.66e+7 & 886 & 3.66e-6 & 6.95e+6 \\
0.30 & 606 & 1.47e-4 & 5.46e+6 & 721 & 8.27e-4 & 1.69e+7 & 886 & 1.09e-5 & 6.88e+6
\enddata
\end{deluxetable*}

Assuming only thermal emission, the combined near-IR and mid-IR spectra provide a strong constraint on the dust temperature and mass, and thus on the IR luminosity.  Assuming optically thin dust with mass $M_{\rm d}$, with a particle radius $a$, at a distance $d$ from the observer, thermally emitting at a single equilibrium temperature $T_{\rm d}$, the flux can be written as \citep[e.g.,][]{hildebrand83}
\begin{equation}
\label{eqn4:flux2}
F_\nu = \frac{M_{\rm d} B_\nu(T_{\rm d}) \kappa_\nu(a)}{d^2},
\end{equation}
where $B_\nu(T_{\rm d})$~is the Planck blackbody function and $\kappa_\nu(a)$ is the dust mass absorption coefficient.

For simple dust populations of a single size composed entirely of either silicate or graphite, the IDL {\tt MPFIT} function \citep{markwardt09} finds the best fit (see Figure \ref{f5}) of Equation \ref{eqn4:flux2} by varying $M_{\rm d}$ and $T_{\rm d}$ to minimize the value of $\chi^2$.  The absorption coefficients, $\kappa$, are given in Figure 4 of \citet{fox10}.  The combined photometry and near-IR spectra of SNe 2006jd and 2005ip are best fit by a multi-component model.  As opposed to a single component with a continuous temperature distribution, a second, hotter component at a single temperature dominates the near-IR flux, while also contributing to the 3.6 \micron~flux.  For the remaining SNe, the small number of photometric data points limits our fits to a single component.  Since a second, hotter component may be contributing to the shorter wavelength flux, we treat the derived dust temperatures for these SNe as upper limits (see \S \ref{sec:echo} below for significance).

Table \ref{tab6} lists the best-fit parameters (i.e., temperature and mass, as well as IR luminosity) for several grain sizes.  Overall, grain size has little consequence on the goodness of fit or the resulting parameters.  The only size-dependent parameter in Equation \ref{eqn4:flux2} is $\kappa$, but Figure 4 of \citet{fox10} shows that the dust-opacity coefficient for both graphite and silicate is independent of grain radius at IR wavelengths ($>$1 \micron) for grain sizes $<$1 \micron, which is typical for most grains.  A grain size of $a = 0.1$~\micron~is therefore assumed throughout the rest of this paper, including the results plotted in Figure \ref{f5}.  

In general, the derived dust temperatures, masses, and luminosities are all higher if we assume silicate grain composition.  The limited number of data points (2 or 3 in most cases), however, limits the best-fitting function in Figure \ref{f5} from distinguishing between the two compositions.  The most obvious differentiating feature is the 9 \micron~silicate emission band, but the warm IRAC data do not extend beyond $\sim$5 \micron.  Both compositions are therefore considered throughout the rest of this paper.  

\begin{figure*}
\epsscale{1}
\begin{center}
\plotone{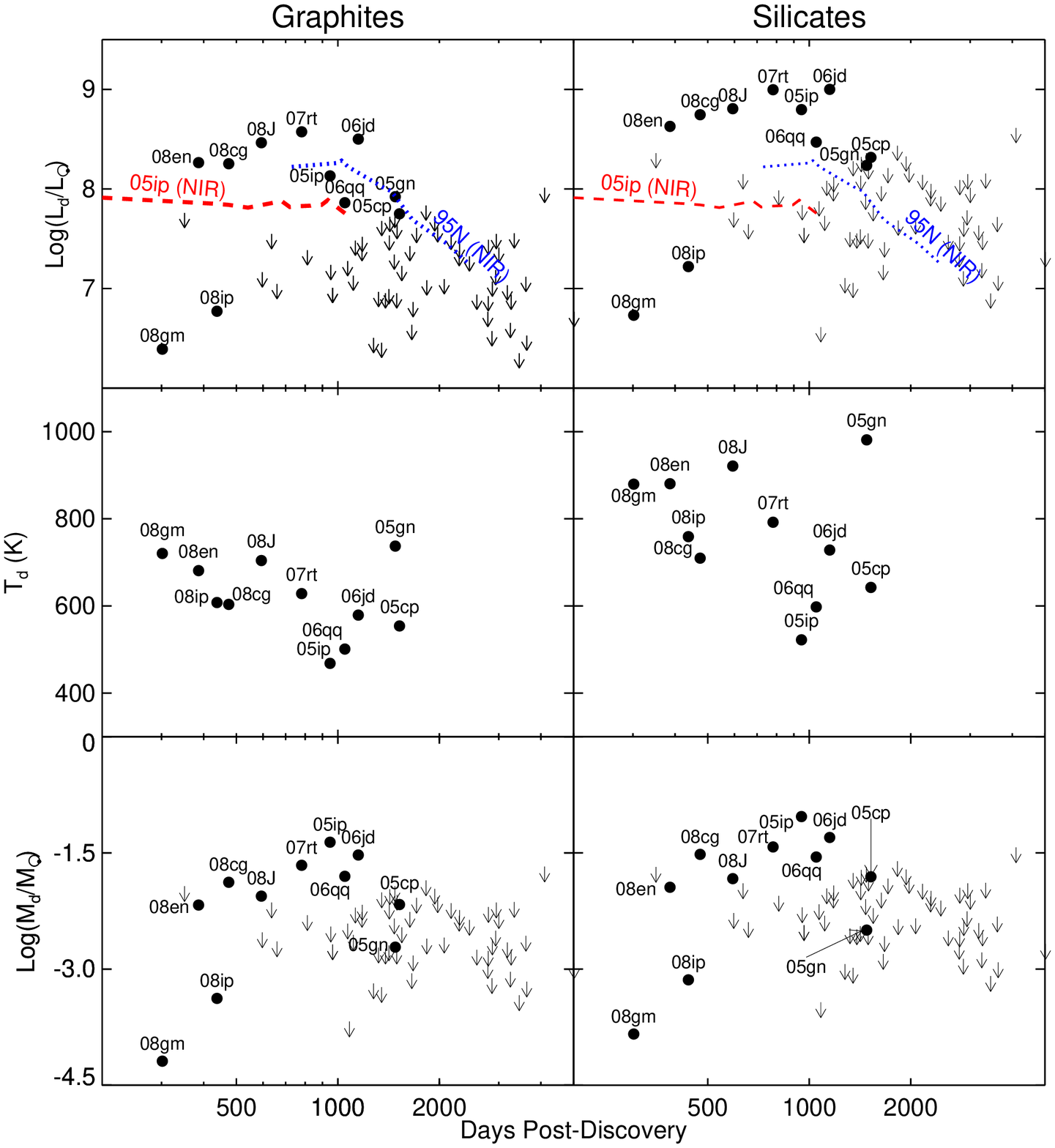}
\caption{The temperature, mass, and IR luminosity of the observed late-time dust component of each SN, as listed in Table \ref{tab6} for $a = 0.1$~\micron~grains of both compositions.  Downward arrows indicate upper limits.  Also plotted is the near-IR evolution of the Type IIn SNe 2005ip \citep{fox09} and 1995N \citep{gerardy02}.  The number of SNe that exhibit late-time IR emission drops dramatically after $\sim 1000$ days, and the trend appears to closely follow the observed decline of SN 1995N.  Generally, the dust masses are $M_{\rm d} > 10^{-3}$~\msolar, and many are even $M_{\rm d} > 10^{-2}$~\msolar.  Ultimately, these variables can be used to disentangle the composition, origin, and heating mechanism of the dust (see \S \ref{sec:source}).  As discussed in \S \ref{sec:model}, many of these results are consistent with LBV progenitors.
}
\label{f6}
\end{center}
\end{figure*}

Figure \ref{f6} plots the temperature, mass, and IR luminosity for $a=0.1$~\micron~grains for both the graphite and silicate fits, as well as the near-IR evolution of the Type IIn SNe 2005ip \citep{fox09} and 1995N \citep{gerardy02}.  The luminosity curve plateaus to a point between 1000 and 2000 days ($\sim$3--5 yr), at which time it seems to slowly fade with an $e$-fold time of several hundred days.  Although many of the points are upper limits, the trend appears to closely follow the observed near-IR decline of SN 1995N.  Furthermore, at least half of the SNe have derived dust masses $M_{\rm d} > 10^{-2}$~\msolar~(and most of the rest have masses $M_{\rm d} > 10^{-3}$~\msolar).  If newly formed, these masses would be some of the largest amounts of such dust observed in SNe to date, providing significant credibility to SN dust models \citep[e.g.,][]{nozawa03,nozawa08}.  Before any conclusions are drawn, however, we must first disentangle the composition, origin, and heating mechanism of the dust.

\section{Analysis: Dust Origin and Heating Mechanism}
\label{sec:source}

The dust can have several different origins and heating mechanisms.  It may be newly formed or it may have been in place at the time of the explosion.  If newly formed, the dust may have condensed in the expanding SN ejecta \citep[e.g.,][]{elmhamdi04} or in the cool, dense shell of post-shocked gas lying between the forward and reverse shocks \citep{pozzo04, smith08jc}.  In both cases, several heating mechanisms are possible, including radioactivity, radiative heating by the optical emission from circumstellar interaction, and collisional heating by hot gas in the reverse shock.  

Alternatively, pre-existing dust may be collisionally heated by hot, shocked gas or radiatively heated by either the peak SN luminosity or the late-time optical emission from circumstellar interaction.  In the radiative heating case, the dust reprocesses the optical light and thermal radiation persists until the dust grains cool sufficiently.  If this pre-existing dust is distributed in a shell with a light-crossing time greater than the duration of the optical emission, an ``IR echo'' is evident due to light-travel-time effects.  Multiple scenarios can also contribute to the late-time IR flux, as in the cases of SNe 2004et \citep{kotak09}, 2004dj \citep{meikle11}, and 2006jc \citep{mattila08}.

With a detailed analysis of SN 2005ip, \citet{fox10} illustrate how to use late-time mid-IR observations, along with optical photometry and spectra, to disentangle the various dust models.  First, a few variables must be defined.  The blackbody radius
\begin{equation}
\label{eqn_rbb}
r_{\rm bb} = \bigg(\frac{L_{\rm bb}}{4 \pi \sigma T_{\rm bb}^4}\bigg)^{1/2},
 \end{equation} 
where $\sigma$~is the Stefan-Boltzmann constant.  The calculation of the blackbody radius assumes an optically thick dust shell.  While our calculations assume the case of optically thin dust, the blackbody radius sets the {\it minimum} shell size of an observed dust component.

The shock radius
\begin{equation}
\label{eqn_rs}
r_{\rm s} = v_{\rm s} t,
\end{equation}
for a constant shock velocity $v_{\rm s}$, defines the maximum radius that the forward shock can travel in a time $t$.  The shock velocity corresponds to the optical spectral line widths.  Three width components are typical for SNe~IIn (see Figures \ref{f7}, \ref{f8}, and \ref{f9}, as well as \S \ref{sec:detections}): narrow ($\lesssim 700$ \kms), intermediate ($\sim 1000$--5000 \kms), and broad ($\sim 5000$--15,000 \kms).  The narrow lines, which give the Type IIn subclass its name, originate in the slow, dense, pre-existing circumstellar environment when excited by X-ray and ultraviolet (UV) emission generated by the forward shock.  The intermediate lines correspond to the decelerated forward-shock front as it passes through the dense clumps in the CSM.  The broad lines, which set the shock velocity, arise from the uninhibited forward shock arising from the rapidly expanding SN ejecta.  Over time, the dense CSM decelerates the entire ejecta.  Typical forward shock velocities in SNe IIn at late times are $v_s \approx 5000-6000$~\kms~\citep[e.g.,][]{smith07, trundle09, rest09}.  

Table \ref{tab7} lists the associated blackbody and shock radii for each SN at the epoch of the $Spitzer$~observation, assuming a shock velocity of $v_s = 5000$~\kms.  Notice that unlike the shock radius, the blackbody radius is {\it not} a function of the observation epoch to the extent that the luminosity is constant.

\begin{deluxetable}{ l c c c c c}
\tablewidth{0pt}
\tabletypesize{\footnotesize}
\tablecaption{Blackbody, Shock, Echo Plateau, and Vaporization Radii \label{tab7}}
\tablecolumns{6}
\tablehead{
\colhead{SN} & \colhead{Epoch} & \colhead{$r_{\rm bb}$\tablenotemark{a}} & \colhead{$r_{\rm s}$ \tablenotemark{b}} & \colhead{$r_{\rm ech}$\tablenotemark{c}} & \colhead{$r_{\rm evap}$\tablenotemark{d}}\\
\colhead{} & \colhead{(days)} & \colhead{(ly)} & \colhead{(ly)} & \colhead{(ly)} & \colhead{(ly)}
}
\startdata
2005cp & 1523 & 0.038 & 0.066 & 2.1 & 0.012 \\
2005gn & 1479 & 0.016 & 0.064 & 2.0 & 0.013\\
2005ip & 948 & 0.031 & 0.040 & 1.3 & 0.01\\
2006jd & 1149 & 0.067 & 0.050 & 1.6 & 0.01\\
2006qq & 1048 & 0.048 & 0.044 & 1.4 & 0.016 \\
2007rt & 780 & 0.057 & 0.034 & 1.1 & 0.02 \\
2008J & 593 & 0.035 & 0.026 & 0.81 & 0.02 \\
2008cg & 474 & 0.036 & 0.020 & 0.65 & 0.02 \\
2008en & 386 & 0.031 & 0.016 & 0.53 & 0.011 \\
2008gm & 301 & 0.004 & 0.014 & 0.41 & 0.0065 \\
2008ip & 438 & 0.008 & 0.020 & 0.60 & 0.015
\enddata
\tablenotetext{a}{The blackbody radius given by Equation \ref{eqn_rbb}.}
\tablenotetext{b}{The shock radius given by Equation \ref{eqn_rs} for $v_s$=5000 \kms.}
\tablenotetext{c}{The IR echo radius given by Equation \ref{eqn_le}.}
\tablenotetext{d}{The vaporization radius is the radius at which the dust temperature equals the dust vaporization temperature.}
\end{deluxetable}

Other useful parameters are the peak supernova UV-optical luminosity, $L_{\rm peak}$, and the late-time optical luminosity, $L_{\rm opt}$.  Dominated by shock heating of the stellar envelope, the peak optical luminosity can vaporize all pre-existing dust within the vaporization radius \citep[e.g.,][]{dwek85},
\begin{equation}
\label{eqn_revap}
r_{\rm evap} = \bigg( \frac{L_{\rm peak}}{16 \pi \sigma T_{\rm evap}^4 \langle Q \rangle} \bigg)^{1/2},
\end{equation}
where T$_{\rm evap}$~is the vaporization temperature of the dust, and $\langle Q \rangle$~is the Planck-averaged value of the dust emissivity.  Figure 8 of \citet{fox10} shows the relationship between the peak luminosity and vaporization radius.  Given the limited number of published observations for each SN during the discovery period, we obtain our best estimates of the peak luminosities from the well-known SN archival website, {\it www.astrosurf.com/snweb2}.  The maximum reported luminosities for each supernova are in the range $3\times10^{8} < L_{\rm peak} < 4\times10^{9}$~\lsolar.  Table \ref{tab7} lists the corresponding vaporization radii, all of which are within $0.0065 \lesssim r_{\rm evap} \lesssim 0.02$~ly.

Whether a dust shell is located at or beyond $r_{\rm evap}$, light-travel-time effects cause the thermal radiation from the dust grains to reach the observer over an extended period of time, thereby forming an ``IR echo'' \citep[e.g.,][see \S \ref{sec:echo} for further discussion]{bode80,dwek83,emmering88}.  The IR echo duration $t_{\rm ech}$ defines the echo radius
\begin{equation}
\label{eqn_le}
r_{\rm ech} = \frac{ct_{\rm ech}}{2}.
\end{equation}
Table \ref{tab7} also lists the associated echo radii for each SN, where $t_{\rm ech}$~is a lower limit determined by the epoch of the latest $Spitzer$~observations.

At later times, once the radioactive component fades, interaction between the reverse shock can generate X-rays and UV radiation \citep[e.g.,][]{draine91,chugai93,chevalier94}.  In the dense CSM, this radiation may be reprocessed and emitted at optical wavelengths, $L_{\rm opt}$.  For example, for both SNe 2005ip \citep{immler07,fox09,smith09ip} and 1995N \citep{gerardy02}, the observed late-time X-ray luminosity is lower than the optical/IR, and for SN 1995N, the UV accessible to the {\it Hubble Space Telescope} has a power comparable to the optical \citep{fransson02}.  While late-time optical observations do not exist for most of the SNe in this paper, the late-time optical observations of SNe 2007rt \citep{trundle09}, 2006jd (see Table \ref{tab5}), and 2005ip \citep{smith09ip} serve as useful references.

\subsection{Newly Formed Dust}
\label{sec:newdust4}

New dust may condense in either the slowly moving ejecta or in the cool, dense shell behind the forward shock (and in front of the reverse shock).  Since the dust will preferentially absorb emission from redshifted (far-side) material, the lack of attenuation in the red wing of the optical emission lines provides a straightforward method for testing whether the dust is newly formed.  (Due to the SN geometry, dust in the slowly moving ejecta closest to the core absorbs redshifted emission from the broad, intermediate, and narrow lines, while dust in the cool dense shell beyond the fastest ejecta absorbs redshifted emission only from the intermediate and narrow components.)

\begin{figure}
\begin{center}
\epsscale{1}
\subfigure{\plotone{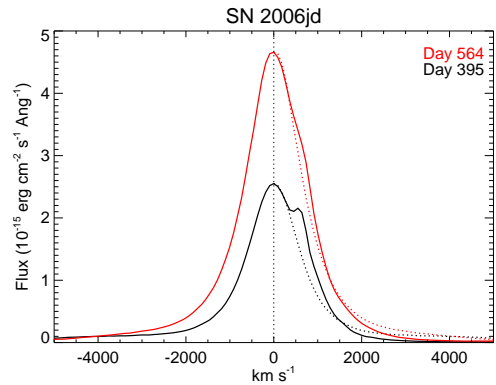}}
\caption{Spectroscopic evolution of the H$\alpha$~line in SN 2006jd at late times.  The underlying continuum has been subtracted.  These spectra show no evidence of absorption.  In fact, they exhibit an excess likely due to the [N~II] $\lambda$6583 line.}
\label{f7}
\end{center}
\end{figure}

\begin{figure}
\epsscale{1.1}
\begin{center}
\plotone{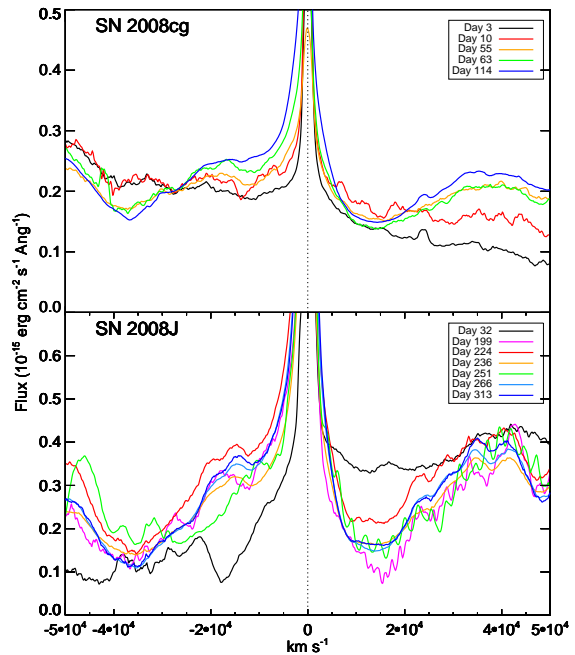}
\caption{Spectroscopic evolution of the broad component of the H$\alpha$~line in SNe 2008cg and 2008J.  An approximate fit of the underlying continuum has been subtracted. It shows increased suppression of the red wing relative to the blue wing with time, especially in SN 2008cg, which offers evidence of new dust formation in the rapidly expanding ejecta.
}
\label{f8}
\end{center}
\end{figure}

\begin{figure}
\epsscale{1.1}
\begin{center}
\plotone{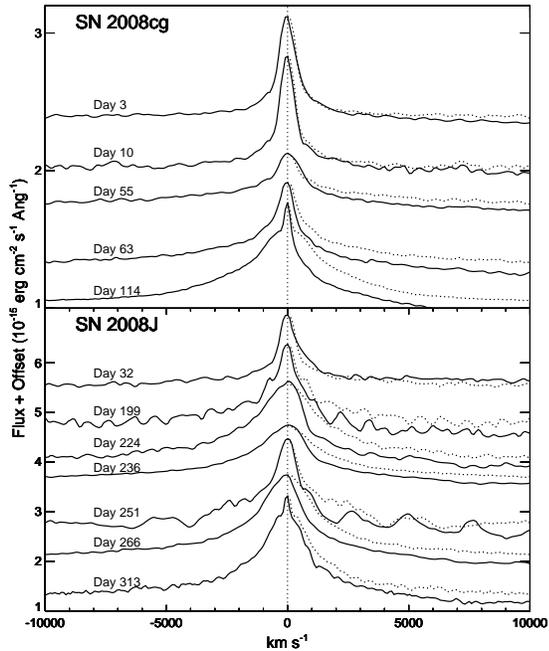}
\caption{Spectroscopic evolution of the intermediate component of the H$\alpha$~line in SNe 2008cg and 2008J.  An approximate fit of the underlying continuum has been subtracted.  The figure reveals an increased suppression of the red side relative to the blue side with time.  The simultaneous absorption in the broad component (Figure \ref{f8}), however, suggests that the dust forms in the ejecta.
}
\label{f9}
\end{center}
\end{figure}

Typically, such emission-line evidence for dust formation occurs at later epochs, once the ejecta expand and become optically thin.  For the first 40--50 days, larger optical depths in the ejecta suppress the redshifted emission \citep{chugai01} so that attenuation in the red wing is little indication of newly formed dust.  \citet{smith09ip} spectroscopically confirm new dust formation in SN 2005ip as early as day 60 post-discovery, similar to the ``Type IIn-like'' SN 2006jc, in which new dust was confirmed as early as 51 days after the peak brightness \citep{smith08jc}.  \citet{trundle09} also provide spectroscopic evidence of dust formation for SN 2007rt.

Of the remaining SNe in this sample, late-time spectra exist only for SNe 2006jd, 2008cg, and 2008J (see Table \ref{tab4}).  Figures \ref{f7}, \ref{f8}, and \ref{f9} plot the H$\alpha$~line from these optical spectra with the underlying continuum subtracted, and the velocity scale chosen with the narrow H$\alpha$~emission feature at $v=0$~\kms.  In some cases, the negative (blueshifted) velocity axis is mirrored (dotted line) on the positive (redshifted) side to provide a qualitative comparison between the redshifted and blueshifted components.  Figure \ref{f7} shows no evidence of absorption in SN 2006jd (in fact, it exhibits an excess likely due to the [N~II] $\lambda$6583 line).

Alternatively, Figures \ref{f8} and \ref{f9} show the broad ($\sim10^4$~\kms) and intermediate ($\sim10^3$~\kms) components of the H$\alpha$~line of SNe 2008cg and 2008J at late times.  Both SNe show some degree of increasing absorption in the red wing relative to the blue wing from days $\ge$63 and 224, respectively.  Simultaneous absorption in both the broad and intermediate components may be evidence for newly formed dust in the ejecta, but other origins exist.  For example, since electron scattering may be responsible for line broadening \citep{chugai01}, it is plausible that this scattering could lead to an asymmetric line profile.  Alternatively, the scattering opacity of dust may produce asymmetric line profiles even when the dust is entirely outside the shock.  Even if the dust were newly formed, however, the masses listed in Table \ref{tab6}  ($M_{\rm d} \le 10^{-2}$~\msolar)~are less than in SN 2005ip and much lower than the $M_{\rm d} = 0.1$--1 \msolar~predicted by the SN dust models \citep[e.g.,][]{nozawa03,nozawa08}.  Furthermore, the relative contribution of this newly formed dust to the overall mid-IR flux is not immediately clear.  Along with the fact that the remaining SNe in this sample do not have late-time spectra, we must therefore consider the pre-existing dust models for all of these SNe.

\subsection{Shock Heating}
\label{sec:shock}

In the shock heating scenario, hot electrons in the post-shocked gas collisionally heat pre-existing dust grains \citep{draine79,draine81,dwek87,dwek08}.  \citet{fox10} derive the {\it estimated}~mass of dust heated by the forward shock, which serves as a useful consistency check when compared to the {\it observed}~dust mass.  For the forward shock at time $t$, the mass of the shocked gas is
\begin{equation}
M_{\rm g} ({\rm M_{\odot}}) \approx 0.28 \bigg(\frac{v_{\rm s}}{{\rm 15,000~\kms}}\bigg)^3 \bigg(\frac{t}{{\rm yr}}\bigg)^2 \bigg(\frac{a}{{\rm \micron}}\bigg)
\label{eqn4:dmr}
\end{equation} 
for a hydrogen atomic mass $m_{\rm H}$, shock velocity $v_{\rm s}$, and grain size $a$, provided that the sputtering time scale $\tau_{\rm sputt} < (t, \tau_{\rm cool})$, where $\tau_{\rm cool}$~is the radiative cooling time scale.  The dust mass is then given by assuming a Galactic dust-to-gas mass ratio, $Z_{\rm d} = M_{\rm d}/M_{\rm g} \approx 0.01$.  

Table \ref{tab8} compares the expected dust masses for the {\it fastest} shocks ($v_s =$ 15,000 \kms) to the observed masses listed in Table \ref{tab6}.  SNe 2008ip, 2008gm, 2005gn, and 2005cp have dust masses consistent with the models, but these cases all assume a forward shock traveling at a constant speed of $v_{\rm s} \approx$ 15,000 \kms~throughout the entire expansion without experiencing any deceleration.  This is not likely for the two events from 2005, but it may be possible for the more recent SNe 2008ip and 2008gm.  The predicted dust masses for the remaining SNe, however, are more than an order of magnitude lower than observed.  Furthermore, the dust-to-gas mass ratio in the stellar winds is likely lower than that of the ISM \citep{williams06, ivezic10}.  A lower ratio would predict even less dust.  These results likely rule out shock heating for most of these events.

\begin{deluxetable}{ l c c c }
\tablewidth{0pt}
\tabletypesize{\footnotesize}
\tablecaption{Predicted Dust Masses From Shock Heating ($v_{\rm s} =$ 15,000 \kms) vs. Observed \label{tab8}}
\tablecolumns{5}
\tablehead{
\colhead{} & \colhead{} & \colhead{Graphites} & \colhead{Silicates}\\
\colhead{SN} & \colhead{Predicted} & \colhead{Observed} & \colhead{Observed}\\
\colhead{} & \colhead{$M_{\rm d}$~(\msolar)} & \colhead{$M_{\rm d}$~(\msolar)} & \colhead{$M_{\rm d}$~(\msolar)}
}
\startdata

2005cp & 4.9e-3 & 6.8e-3 & 1.6e-2 \\
2005gn & 4.6e-3 & 1.9e-3 & 3.2e-3 \\
2005ip & 1.9e-3 & 4.3e-2 & 9.3e-2 \\
2006jd & 2.8e-3 & 3.0e-2 & 5.0e-2 \\
2006qq & 2.3e-3 & 1.6e-2 & 2.8e-2 \\
2007rt & 1.3e-3 & 2.2e-2 & 3.8e-2 \\
2008J & 7.4e-4 & 8.7e-3 & 1.5e-2 \\
2008cg & 4.7e-4 & 1.3e-2 & 3.0e-2 \\
2008en & 3.1e-4 & 6.7e-3 & 1.1e-2 \\
2008gm & 1.9e-4 & 6.5e-5 & 1.4e-4 \\
2008ip & 4.0e-4 & 4.2e-4 & 7.3e-4'
\enddata
\end{deluxetable}

\subsection{Radiative Heating}
\label{sec:echo}

Optical emission from the SN may radiatively heat a shell of pre-existing dust at a radius $r$ to a temperature $T_{\rm d}$.  \citet{fox10} outline three basic scenarios.  For the first two scenarios, the peak SN luminosity heats the dust and forms an IR echo \citep[e.g.,][]{wright80}. In the first scenario, the dust may be distributed spherically symmetrically around the star at the time of the explosion (most likely formed by a steady wind from the progenitor).  The SN peak luminosity then vaporizes the dust out to the vaporization radius $r_{\rm evap}$, and heats the inside of the remaining dust shell to a temperature roughly equal to the vaporization temperature ($T_{\rm d} \approx T_{\rm evap}$).  Alternatively, the dust may be distributed in a shell at a radius $r > r_{\rm evap}$ (most likely formed by a progenitor eruption hundreds to thousands of years prior to the SN).  The SN peak luminosity heats the inside of this shell to a temperature $T_{\rm d} \le T_{\rm evap}$, where the relationship between the two temperatures is generally given as
\begin{equation}
\label{eqn_tdust}
T_{\rm d} = T_{\rm evap}\bigg(\frac{L_{\rm opt}}{L_{\rm evap}}\bigg)^{\frac{1}{4}},
\end{equation}
where $L_{\rm opt}$~is the observed SN optical luminosity and $L_{\rm evap}$~is the luminosity required to vaporize the dust at the given radius.  In many cases \citep[e.g.,][]{fox09}, the resulting IR luminosity echo plateaus on year-long time scales, corresponding to the light travel time across the inner edge of the dust shell, where the shell radius is given by Equation \ref{eqn_le}.

In the final scenario, the dust may be distributed in a shell at radii between the vaporization and echo radii, $r_{\rm evap} \le r \le r_{\rm ech}$ (formed by either a steady wind or progenitor eruption).  The late-time optical luminosity generated by circumstellar interaction, $L_{\rm opt}$, continuously heats the dust shell.  This scenario is not so much a traditional IR echo as it is a reprocessing of the optical emission by the dust.\footnote{Some authors refer to this as a circumstellar interaction echo \citep{gerardy02}, but the reader should note that the heating mechanism is radiation from the shocks, rather than direct collisional heating by the shocks themselves.}  If the circumstellar interaction occurs on a time scale greater than the light travel time across the dust shell (which we assume it does in these cases), the shell radius does not set the IR echo plateau length.  The observed flux therefore accounts for the entire shell.

For a simple light-echo model dominated by flux from only the innermost, warmest dust at a single temperature, $T_{\rm d}$, balancing the energy absorbed and emitted by the dust grains gives the relationship between the SN optical luminosity and the dust temperature and inner shell radius as \citep{fox10}
\begin{equation}
\label{eqn4:lbol}
L_{\rm opt}  = \frac{64}{3} \rho a r^2 \sigma T_{\rm SN}^4 \frac{\int{B_\nu (T_{\rm d}) \kappa(\nu) d\nu}}{\int{B_\nu(T_{\rm SN}) Q_{\rm abs}(\nu) d\nu}}, 
\end{equation}
for a dust bulk (volume) density $\rho$ and an effective SN blackbody temperature $T_{\rm SN}$.  When using this equation, we generally assume $a=0.1$~\micron~grains and a SN temperature $T_{\rm SN} \approx$ 10,000 K, although the results are fairly insensitive to the choice of temperature.  The luminosity is treated as a central point source, assuming the emitting region is internal to a spherically symmetric dust shell.

\begin{figure*}
\epsscale{0.57}
\begin{center}
\subfigure[Infrared Echo for $r=r_{\rm ech}>r_{\rm evap}$]{\label{f10a} \plotone{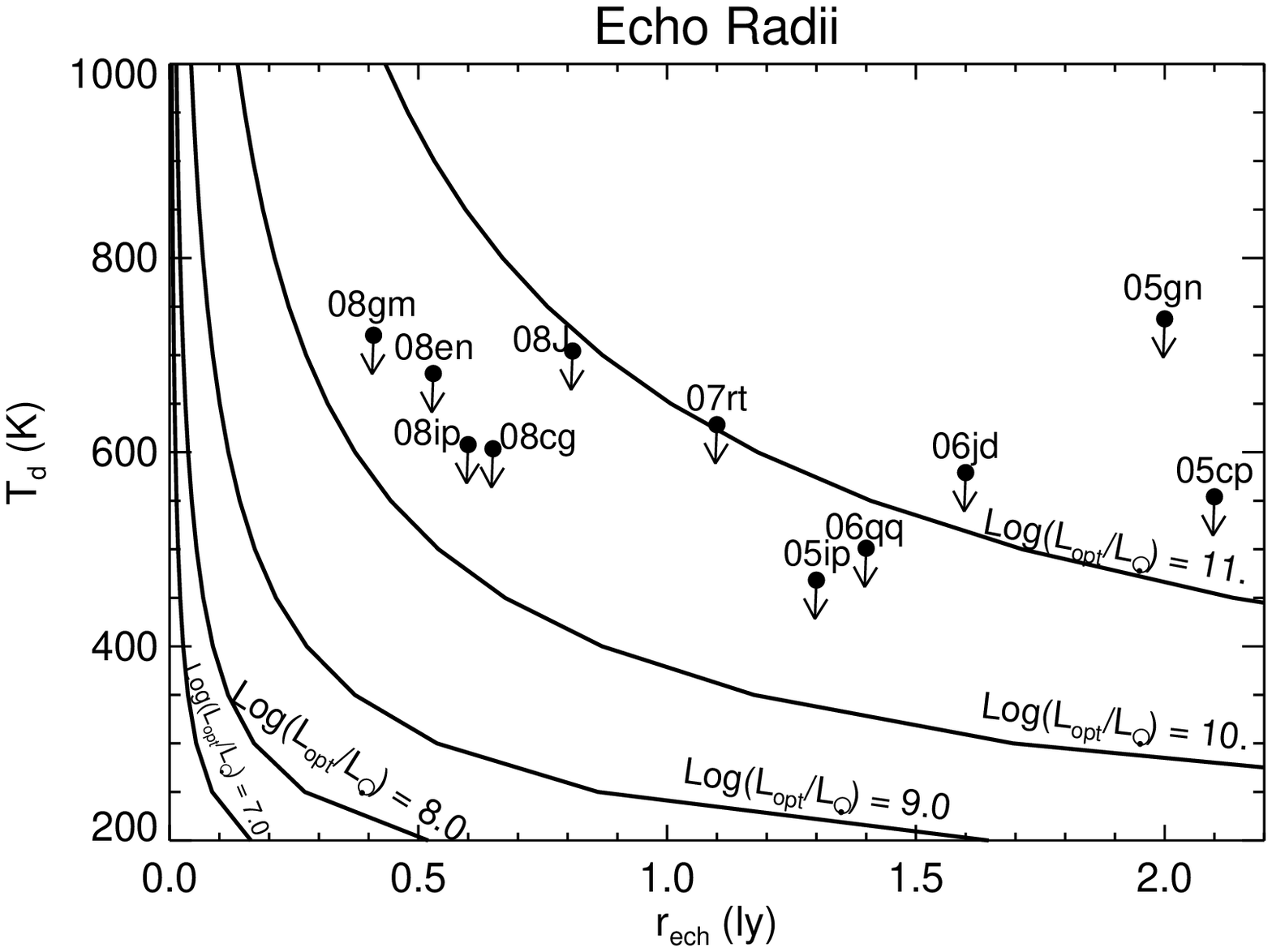}}
\subfigure[Continuous Radiative Heating of Dust Shell for $r_{\rm evap} \le r_{\rm bb} < r_{\rm ech}$]{\label{f10b} \plotone{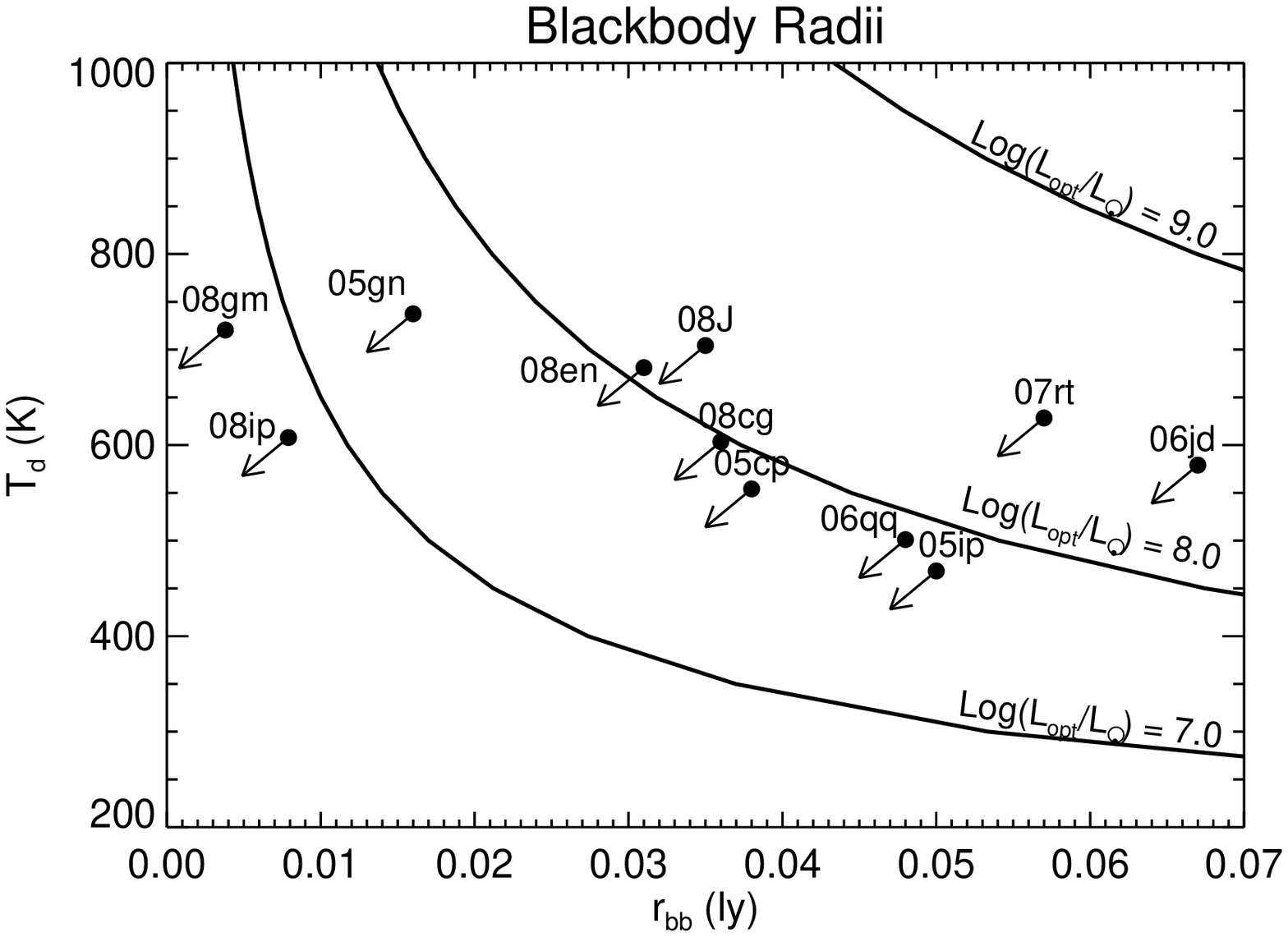}}
\caption{Testing the possibility of radiative heating of a pre-existing dust shell (only graphites are considered here).  The dust temperature, T$_{\rm d}$, is given as a function of radius, $r$, for contours of constant optical luminosity.  Overplotted are the {\it observed} dust temperatures and radii.  The arrows signify that these values are all upper limits given the lack of constraints in our fitting routines.  In case (a), the dust shell lies at the echo radius $r_{\rm ech}$, defined by the observed light-curve duration, and is heated by the peak SN luminosity, which is typically in the range $10^9 \le L_{\rm opt} < 10^{10}$~\lsolar.  The required luminosities to heat the dust are too high.  In case (b), the dust lies at the minimum radius given by $r_{\rm bb}$ and is heated by optical emission generated by continuous circumstellar interaction, typically observed to be $10^7 \le L_{\rm opt} \le 10^{8}$~\lsolar.  Aside from a few caveats discussed in the text, the observed temperatures are consistent with the expected values given by the contours.
}
\label{f10}
\end{center}
\end{figure*}

Using Equation \ref{eqn4:lbol}, Figure \ref{f10} plots contours of the {\it expected} dust temperature as a function of shell radius $r$ for various optical luminosities.  We choose luminosities $10^7 \le L_{\rm opt} \le 10^{11}$~\lsolar, which span the observed peak and late-time optical luminosities for these SNe.  Overplotted  are the {\it observed} dust temperatures and radii derived from the $Spitzer$~data (see Tables \ref{tab6} and \ref{tab7}).  Only dust with graphite composition is considered, as silicate grains require much higher luminosities than observed ($>10^{11}$ \lsolar) to heat dust grains of comparable sizes to the observed temperatures.  

Considering the first scenario above, the SN peak luminosity vaporizes all dust out to the vaporization radius, $r_{\rm evap}$, which is defined by the radius at which the dust temperature equals the dust vaporization temperature (for graphite, $T_{\rm evap} \approx$ 2000~K).  None of the observed dust temperatures in Table \ref{tab6} even approach the vaporization temperature of graphite.  Furthermore, we can rule out this scenario by comparing the expected vaporization radii to the observed duration of the IR echos.  Table \ref{tab7} shows that the vaporization radii are all $r_{\rm evap} \lesssim 0.02$~ly, which is too small to produce an IR echo over the observed year-long time scales.  Of course, the unobserved shock breakout in the minutes to hours following the supernova can approach peak values $\sim 10^{11}$~\lsolar~\citep{soderberg08,nakar10}.  Figure 8b in \citet{fox10} shows that such values can vaporize 0.1 \micron~dust grains out to $r_{\rm evap} \approx 0.2$~ly.  Still, these radii are too small to produce an IR echo over the observed year-long time scales.

Figure \ref{f10a} tests the second scenario by comparing the expected and observed temperatures assuming the dust lies at the echo radii $r_{\rm ech}$ defined by the IR echo plateau duration $t_{\rm ech}$.  At these radii, the SNe require large peak luminosities ($>10^{10}$~\lsolar) to power the observed echo.  Not only are the required peak luminosities significantly greater than the measured ones, but they are larger than those of the most luminous core-collapse events ever found \citep{quimby07}.  The unobserved shock breakout luminosities are indeed higher, but the breakout typically lasts only for about 30 minutes, even in the case of an extended red supergiant progenitor.  While the shock breakout luminosity may be large enough to heat the dust to the observed temperature, the total radiated energy from the breakout ($E_{\rm breakout} \approx 10^{11}~$\lsolar~$\times$~30 min $< 10^{48}$~erg) is not nearly enough to power the observed echo ($E_{\rm echo} \approx 10^{8.5}~$\lsolar~$\times$~1000 d $\approx 10^{50}$~erg), assuming an IR luminosity $L_{\rm IR} \approx 10^{8.5}$~\lsolar~extending throughout the 1000 day period (see Figure \ref{f6}).  Even if we take an echo lasting for only 100 days, the total radiated IR energy is still too large to have been powered by the shock breakout.  This fact makes unlikely any IR echo powered by the peak SN luminosity from shock breakout.

Figure \ref{f10b} tests the final scenario, in which the optical luminosity generated by continuous interaction between the forward shock and CSM offers an alternative heating mechanism.  In this case, the dust is assumed to lie at the smallest radius allowed by the observations, defined by the blackbody radius, $r_{\rm bb}$.

The observed dust temperatures for most of the SNe require optical luminosities $10^7 <L_{\rm opt}<10^8$~\lsolar.  While such late-time optical luminosities are larger than observed in most SN subclasses, they are quite realistic for Type IIn events.  For example, late-time observations show $L_{\rm opt}$ (SN 2005ip) = $10^{7.8}$ \lsolar~on days $>$900 \citep{smith09ip}, $L_{\rm opt}$ (SN 2006jd) = $10^{7.7}$ \lsolar~on day 1609 (Table \ref{tab5}),  and $L_{\rm opt}$ (SN 2007rt) $\approx 10^{8}$ \lsolar~on day 562 \citep{trundle09}\footnote{We reconcile the fact that our most recent optical observations do not detect SN 2007rt by noting that they were made nearly 500 days following the original $Spitzer$~observations.}.

The dust temperatures for some of the SNe, particularly SNe 2007rt and 2006jd, do fall above the expected values for a luminosity $L_{\rm opt} \approx 10^{8}$~\lsolar.  However, in these cases a second, hotter dust component may be contributing to the mid-IR flux.  Already, we know from the fits in Figure \ref{f5} that SN 2006jd has a hotter dust component.  Also, \citet{trundle09} show that new dust forms in SN 2007rt, although they do not measure the temperature.  Similarly, Figures \ref{f8} and \ref{f9} exhibit evidence for new dust formation in SNe 2008cg and 2008J.  As in SN 2005ip \citep{fox10}, this newly formed dust may be hotter than the mid-IR dust component.  In these cases, the observed $Spitzer$~temperatures would be only upper limits.  The resulting lower dust temperatures and corresponding lower blackbody radii would place these points significantly closer to their luminosity contour lines.  

Late-time optical observations are not available for SNe 2005cp, 2005gn, 2006qq, 2008J, 2008cg, 2008en, 2008gm, and 2008ip, but almost all of these points fall close to or along luminosity contours comparable to observed late-time optical luminosities of SNe 2005ip, 2006jd, and 2007rt.  In general, the data are fairly consistent with heating by optical emission generated by continuous circumstellar interaction.  \citet{gerardy02} reached a similar conclusion in their study of late-time emission from five SNe.  Two of the Gerardy et al. targets, SNe 1999Z and 1997ab, are actually on our initial target list.  The fact that we failed to detect any late-time mid-IR emission comes as little surprise given that the forward shock would eventually destroy the pre-existing dust shells, a trend we observe in Figure \ref{f6} and discuss more below in \S \ref{sec:model}.

A simple calculation shows that the pre-existing dust shells in this model are optically thin at IR wavelengths, consistent with assumptions for deriving the dust mass in Equation \ref{eqn4:flux2}.  For a dust shell given by a blackbody radius $r_{\rm bb}$, the optical depth can be written as
\begin{equation}
\tau_{\rm bb} = \frac{M_{\rm d}}{4 \pi r_{\rm bb}^2}  \kappa_{\rm avg},
\label{eqn4:tau}
\end{equation}
where $\kappa_{\rm avg} = 435$~cm$^2$~g$^{-1}$~is the absorption coefficient for graphite averaged over 1--15 \micron~(see Figure 4 of \citealt{fox10}).  For each SN except SN 2005ip, $\tau_{\rm bb} < 1$.  Using the additional spectroscopic data for SN 2005ip provided by \citet{fox10}, however, we derive an optical depth $\tau_{\rm bb} \approx 0.5$.  Furthermore, since the blackbody radii are only lower limits, these optical depths are all upper limits.

\section{Trends, Progenitors, and Overall Model}
\label{sec:model}

For the above scenario, in which UV and optical emission from circumstellar interaction radiatively heats pre-existing dust, the observed dust mass accounts for the entire dust shell since the shell-radius light travel time ($2r_{\rm bb}/c$) is significantly smaller than the circumstellar interaction time scale given by the epoch of observation (see Table \ref{tab7}).  Assuming a dust-to-gas mass ratio expected in the H-rich envelope of a massive star, $Z_{\rm d} = M_{\rm d}/M_{\rm g} \approx 0.01$, the pre-existing dust mass corresponds to a progenitor mass-loss rate 
\begin{eqnarray}
\label{eqn:ml}
\mdot & = & \frac{M_{\rm d}}{Z_{\rm d} \Delta r} v_{\rm w} \\ 
& = & 0.75 \Big(\frac{M_{\rm d}}{\rm M_{\odot}}\Big) \Big(\frac{v_{\rm w}}{120~\rm km~s^{-1}}\Big) \Big(\frac{0.05~\rm ly}{r}\Big) \Big(\frac{r}{\Delta r}\Big) \frac{{\rm M_{\odot}}}{\rm yr},
\end{eqnarray}
for a wind speed $v_{\rm w}$.  The SNe in \S \ref{sec:detections} have narrow line profiles with FWHM in the range $\sim 100$--500 \kms.  Since these narrow lines originate in the slow, dense, pre-existing circumstellar environment, they are typically assumed to directly correspond to the progenitor wind speed.  Assuming a thin shell, ${\Delta r}/r = 1/10$, Table \ref{tab9} lists the associated mass-loss rate for each SN.

\begin{deluxetable}{ l c c c}
\tablewidth{0pt}
\tablecaption{Mass-loss Associated with Radiatively Heated Graphite Dust Shell \label{tab9}}
\tablecolumns{4}
\tablehead{
\multirow{2}{*}{SN} & \colhead{$\dot{M} \times (10\Delta r)/r$} & $t_{\rm eruption}$ & $L_{\rm opt}$\tablenotemark{a}\\
 &  (\msolar~yr$^{-1}$) & (yr) & log (L / L$_{\odot}$)
}
\startdata
2005cp & 2.2e-3 & 30 & 7.45\\
2005gn & 7.5e-4 & 25 & 7.10\\
2005ip & 1.8e-3 & 235 & 7.88\\
2006jd & 2.8e-3 & 106 & 7.68\\
2006qq & 9.2e-3 & 17 & 7.72\\
2007rt & 1.2e-3 & 181 & 7.79\\
2008J & 1.6e-3 & 55 & 7.43\\
2008cg & 2.3e-3 & 57 & 7.59\\
2008en & 1.4e-3 & 49 & 7.38\\
2008gm & 1.6e-4 & 4 & 6.43\\
2008ip & 3.3e-4 & 13 & 6.75
\enddata
\tablenotetext{a}{The expected optical luminosity generated by the forward shock, given by Equation \ref{eqn_lumloss}.}
\end{deluxetable}

The properties in Table \ref{tab9} are generally consistent with LBVs during their giant eruptions, which have wind speeds on the order of hundreds of \kms~and mass-loss rates $\dot{M} = 10^{-5} - 10^{-4}$ \ml~\citep{humphreys94}, but can reach much higher \citep{smith06a,smith07}.  In fact, the total dust-shell masses are in the range $10^{-5} < M_{\rm d} < 10^{-1}$~\msolar~(see Table \ref{tab6}), which, assuming a dust-to-gas mass ratio $Z_{\rm d} \approx 0.01$, are also consistent with the range of LBV shell masses observed by \citet{smith06a}.  A LBV has already been directly identified as the likely progenitor of SN 2005gl \citep{gal-yam09}, and, indirectly, LBVs have emerged as the likely progenitors of many other SNe IIn \citep[e.g.,][]{chugai94, salamanca02,gal-yam07,smith07,smith08tf, smith09gy, kiewe10}.

Table \ref{tab7} highlights the fact that in almost every case (SNe 2008gm and 2008ip aside), the blackbody radius is larger than the vaporization radius by nearly a factor of two ($r_{\rm bb} \ge 2 r_{\rm evap}$).  If the dust were associated with a steady wind, the resulting shell size would have been more comparable to the vaporization radius.  Instead, these results suggest that the progenitor may have undergone a period of increased mass-loss prior to the SN at a time $t_{\rm eruption} \approx r_{\rm bb}/v_{\rm w}$, which corresponds to tens to hundreds of years (see Table \ref{tab9}).  Such an eruptive event is inferred in many other Type IIn events \citep[e.g.,][]{chugai94, salamanca02, gal-yam07,smith07, smith08tf, smith09gy, kiewe10}.

Alternatively, Figure 8b in \citet{fox10} shows that an unobserved shock breakout peak luminosity of only $L_{\rm peak} \approx 10^{10}$~\lsolar~\citep{soderberg08,nakar10} can vaporize 0.1 \micron~dust grains out to $r_{\rm evap} \approx 0.05$~ly, which is much more within the range of the observed blackbody radii.  In this case, there is little evidence for an eruptive event.  Instead, the progenitor more likely underwent a period of extreme, but steady, mass-loss prior to the SN, thereby enshrouding the progenitor in a massive gas and dust shell.

To test each possibility, we can compare the observed late-time optical luminosities to the expected values for various CSM densities.  We use the expression \citep[e.g.,][]{chugai92} to write the expected shock luminosity as a function of progenitor wind density, $w=\mdot/v_w$,
\begin{equation}
\label{eqn_lumloss}
L_{\rm opt} = w \frac{\epsilon v_{s}^3}{2},
\end{equation}
where $v_w$~is the preshock wind speed and $\epsilon < 1$~is the efficiency of converting shock kinetic energy into visual light.  While the conversion efficiency varies greatly depending on shock speed and wind density, we assume a value $\epsilon \approx 0.5$, acknowledging this value may be high.  The preshock wind speeds are given by the narrow line widths reported in \S \ref{sec:detections} (and $v_w \approx 300$~\kms~for unreported lines).  

Table \ref{tab9} lists the expected optical luminosity produced by a $v_{\rm s} \approx 5000$~\kms~forward shock moving through the CSM assuming a mass-loss rate equal to the value derived above for each dust shell.  Except for SNe 2008ip and 2008gm, the predicted values are nearly equal to the observed late-time optical luminosities reported above for SNe 2005ip, 2006jd, and 2007rt.  This consistency suggests the wind density throughout the CSM is roughly equal to the value calculated for the dust shell, thereby favoring the extreme and steady wind scenario above.     

The story starts to get interesting when we revisit Figure \ref{f6}.  Overall, a significant percentage of the SNe~IIn from the past 10 years ($\sim$15\%) show late-time mid-IR emission (and the percentage is over 50\% when only considering SNe with explosion dates within 5 years of the observations.)  After about 1000--2000 days (3--5 years) post-discovery, the luminosity begins to steadily decline.  This decline may occur because the forward shock ultimately overruns and destroys the dust shell.  Alternatively, the radiative efficiency of the shock will drop when the cooling time becomes longer than the expansion time.  In either case, the apparent consistency amongst the detected SNe suggests that in these instances, many of the progenitors may have experienced similar mass-loss histories, although additional epochs of data will be necessary to confirm this possibility.  If indeed the progenitors are LBVs, which are associated with initial stellar masses $M_0 > 25$~\msolar\ \citep{smith04}, this result could have significant implications on mass-loss properties of the high mass, low-metallicity stars observed in the early universe \citep[e.g.,][]{smith06a}.  These conclusions are also particularly noteworthy for stellar evolution theorists, whose models do not place LBVs in the final pre-SN phase \citep{schaller92, langer93, langer94, stothers96, maeder05, maeder08}.  

Of course, LBVs are not necessarily the progenitors of all SNe~IIn, and several caveats should be mentioned.  Primarily, there is a significant diversity among the targets in this sample.  Not only can some of the SNe with late-time emission be explained with other models (e.g., shock heating), but a large percentage of the sample does not exhibit any late-time X-ray, optical, IR, or radio emission \citep{dyk96}.  Of well-studied SNe~IIn, many show varying degrees of circumstellar interaction \citep{li02,trundle09}.  In the extreme case of SN 1986J, for example, late-time optical emission is observed more than two decades post-detection \citep{milisavljevic08}.  (The faded H$\alpha$~line relative to the strong [O I] line, however, suggests that the bulk of the emission is generated by the reverse shock, {\it not}~the forward shock.)  Not only do the observations vary, but several other SNe~IIn have shown evidence for red supergiant (RSG) progenitors \citep{fransson02,smith09rsg}.  Furthermore, other authors invoke several additional scenarios to explain the dense circumstellar shells, including wind-blown bubbles, high-density clumps, very short Wolf-Rayet stages following the LBV outbursts \citep{dwarkadas11}, mass ejection caused by frictional heating of a common envelope in a binary system \citep{tutukov92}, and even relic disks of proto-stellar material left over from the star formation \citep{metzger10}.

\section{Conclusion}
\label{sec:con}

The $Spitzer$~survey in this paper targets 68 SNe~IIn discovered in 1999--2008 without late-time mid-IR observations, as well as the well-studied Type IIn SN 1997ab.  Ten of the 69 SNe ($\sim15$\%) display late-time mid-IR emission from warm dust.  The late-time optical and infrared observations presented here constrain the various origins and heating mechanisms of the warm dust, which are summarized by Table \ref{tab10}.  Although some SNe show evidence of newly formed dust or shock heating of pre-existing dust, observations for almost every SN are consistent with the scenario in which pre-existing dust is radiatively heated by optical emission from late-time circumstellar interaction between the forward shock and dense CSM (see the third scenario of \S \ref{sec:echo}).  A similar conclusion was reached by \citet{gerardy02} regarding several other Type IIn events that have since faded.

Although the model has not yet been confirmed, a trend amongst the Type IIn subclass begins to emerge.  The late-time IR emission associated with this subclass appears to originate from pre-existing dust expelled by the progenitor star, which ultimately forms a shell when the interior dust is vaporized by the SN shock breakout.  With some basic assumptions, the dust-shell properties indicate that the progenitor winds and mass-loss rates are most consistent with the properties of LBVs.  In almost every case, the data point to extreme, but steady, mass-loss histories as the progenitors evolve along toward their stellar death.  These results therefore contribute to the growing evidence that SNe are not likely major sources of dust, although their progenitors might be if the forward shock does not ultimately destroy the dust.

\begin{deluxetable}{ l l l l l }
\tablewidth{0pt}
\tablecaption{Evidence for Dust Origin and Heating Mechanism\label{tab10}}
\tablecolumns{3}
\tablehead{
\colhead{SN} & \colhead{Newly} & \colhead{Shock} & \colhead{IR} & \colhead{Shock}\\
& \colhead{Formed?} & \colhead{Heating?} & \colhead{Echo?} & \colhead{Echo?}  
}
\startdata
2005cp & no & maybe & no & yes\\ 
2005gn & no & maybe & no & maybe\\ 
2005ip & yes & no & no & yes\\ 
2006jd & no & no & no & maybe\\ 
2006qq & no & no & no & yes\\ 
2007rt & yes\tablenotemark{a} & no & no & maybe\\ 
2008J & maybe & maybe & no & maybe\\ 
2008cg & yes & no & no & yes\\ 
2008en & no & maybe & no & yes\\ 
2008gm & no & yes & no & yes\\
2008ip & no & yes & no & yes
\enddata
\tablenotetext{a}{\citet{trundle09}.}
\end{deluxetable}

Still, several caveats exist that offer alternative explanations for the dense dust shells without invoking LBV progenitors.  The advantage to the model in this paper is that it allows us make predictions.  If the dust is distributed in a shell, and the heating mechanism is indeed continuous radiation generated by the forward shock, we expect the shock to eventually overtake the shell radius and destroy the dust.  As the dust is destroyed, the late-time mid-IR emission should begin to significantly decrease.  While we see this general trend in Figure \ref{f6}, no single SN~IIn has multi-epoch mid-IR data that reveal the SN ``turning off.''

We can calculate the approximate point in time when this will happen as $t \approx r_{\rm bb}/v_{\rm s}$, where $v_{\rm s}$~is the forward-shock velocity.  For the observed blackbody radii, $r_{\rm bb} < 0.1$ ly, and a shock velocity $v_{\rm s} \approx 5000$ \kms, we would expect the shock to overtake the shell at a time $t < 5$ yr.  Once the shock destroys the dust, the luminosity decline will occur on time scales consistent with the dust-shell radius ($t_{\rm decline} = 2 r_{\rm bb}/c \approx 3$~yr), as the dust destruction on the near side of the shell will be observed before dust destruction on the far side.  This time scale is consistent with the apparent decline (determined from the upper limits) in Figure \ref{f6}.  Of course, the blackbody radius assumes an optically thick shell and only sets a lower limit.  The actual dust radii are likely much larger.  In addition, slower shock speeds or asymmetric shell geometries would result in varying time scales, but mid-IR monitoring of these SNe will help to constrain the various models and geometries.

\vspace{10 mm}

This work is based on observations made with the {\it Spitzer Space Telescope} (PID 60122), which is operated by the Jet Propulsion Laboratory, California Institute of Technology, under a contract with NASA. Support for this work was provided by NASA through an award issued by JPL/Caltech.  O.D.F. is grateful for support from the NASA Postdoctoral Program (NPP).  R.A.C. was supported by NSF grant AST--0807727. A.V.F. is grateful for the support of NSF grant AST--0908886 and the TABASGO Foundation. J.M.S. thanks Marc J. Staley for a graduate fellowship. Some of the data presented herein were obtained at the W. M. Keck Observatory, which is operated as a scientific partnership among the California Institute of Technology, the University of California, and NASA; the observatory was made possible by the generous financial support of the W. M. Keck Foundation. The Kast spectrograph on the Lick 3-m Shane telescope was funded by a gift from Bill and Marina Kast. We thank the staffs of the Lick and Keck Observatories for their assistance with the observations. We are also grateful to many students and postdocs who helped take and reduce the optical spectra.


\bibliographystyle{apj}
\bibliography{references2}

\end{document}